\documentclass[journal]{IEEEtran}
\usepackage{graphicx}
\usepackage{amssymb}
\usepackage{amsmath}
\usepackage{amsthm}
\usepackage{array}
\usepackage{multirow}
\usepackage{rotating}
\usepackage[ruled]{algorithm2e}
\usepackage{caption}
\usepackage[numbers,sort&compress]{natbib}
\usepackage{color}
\usepackage{hyperref}

\theoremstyle{plain}

\theoremstyle{definition}

\theoremstyle{remark}

\newcommand{\B}[1]{\mathbb{#1}}
\newcommand{\C}[1]{\mathcal{#1}}

\makeatletter

\newcommand{\Rmnum}[1]{\expandafter\@slowromancap\romannumeral #1@}
\makeatother

\begin{document}

\title{Testing Scenario Library Generation for Connected and Automated Vehicles: An Adaptive Framework}

\author{Shuo~Feng,
	Yiheng~Feng,
	Haowei~Sun,
	Yi~Zhang,~\IEEEmembership{Member,~IEEE},
	and Henry~X.~Liu,~\IEEEmembership{Member,~IEEE}
	
	\thanks{This research was partially funded by US Department of Transportation (USDOT) Region 5 University Transportation Center: Center for Connected and Automated Transportation (CCAT) at the University of Michigan, Ann Arbor.  \emph{(Corresponding author: Henry X. Liu)}
	}
	\thanks{Shuo Feng is with the Department of Automation, Tsinghua University, Beijing 100084, China, and also with the Department of Civil and Environmental Engineering, University of Michigan, Ann Arbor, MI, 48109, USA. (e-mail: s-feng14@mails.tsinghua.edu.cn)}
	
	\thanks{Yiheng Feng is with the University of Michigan Transportation Research Institute, 2901 Baxer Rd, Ann Arbor, MI, 48109, USA. (e-mail: yhfeng@umich.edu)}
	
	\thanks{Haowei Sun and Henry X. Liu are with the Department of Civil and Environmental Engineering, University of Michigan, Ann Arbor, MI, 48109, USA.  (e-mail: haoweis; henryliu@umich.edu)}
	
	\thanks{Yi Zhang is with the Department of Automation, Tsinghua University, Beijing 100084, China. (e-mail: zhyi@mail.tsinghua.edu.cn).}
}

%

\maketitle

\begin{abstract}
How to generate testing scenario libraries for connected and automated vehicles (CAVs) is a major challenge faced by the industry. In previous studies, to evaluate maneuver challenge of a scenario, surrogate models (SMs) are often used without explicit knowledge of the CAV under test. However, performance dissimilarities between the SM and the CAV under test usually exist, and it can lead to the generation of suboptimal scenario libraries. In this paper, an adaptive testing scenario library generation (ATSLG) method is proposed to solve this problem. A customized testing scenario library for a specific CAV model is generated through an adaptive process. To compensate for the performance dissimilarities and leverage each test of the CAV, Bayesian optimization techniques are applied with classification-based Gaussian Process Regression and a newly designed acquisition function. Comparing with a pre-determined library, a CAV can be tested and evaluated in a more efficient manner with the customized library. To validate the proposed method, a cut-in case study is investigated and the results demonstrate that the proposed method can further accelerate the evaluation process by a few orders of magnitude.
\end{abstract}


\IEEEpeerreviewmaketitle

\section{Introduction}

\IEEEPARstart{T}{esting} scenario library generation (TSLG) is a major challenge in evaluating connected and automated vehicles (CAVs). A scenario describes the temporal development in a sequence of scenes, where a scene is a snapshot of the environment including stationary elements (e.g., road geometry) and dynamic elements (e.g., background vehicles) \cite{ulbrich2015defining}. Given an operational design domain (ODD) \cite{J3016_201806}, there could exist millions of scenarios with different parameters, e.g., different maneuvers of background vehicles. A testing scenario library is defined as a critical subset of scenarios that can be used for the evaluation of certain performance metrics (e.g., safety). In the past few years, increasing research efforts have been made to solve the TSLG problem \cite{jung2007worst, PEGASES, li2016intelligence, zhao2017accelerated, zhao2018accelerated, li2018artificial, mullins2018adaptive, zhang2018accelerated, li2019parallel} (see \cite{feng2019testingI} and references therein). 
However, most existing methods have limitations in either scenario types that can be handled, CAV models that can be applied, or performance metrics that can be evaluated. 

To overcome these limitations, a systematic framework was proposed in our previous studies \cite{feng2019testingI, feng2019testingII, feng2020safety}. Each testing scenario was evaluated by a newly proposed measure, scenario criticality, which can be computed as a combination of exposure frequency and maneuver challenge. The exposure frequency can be obtained by using naturalistic driving data (NDD). To evaluate the maneuver challenge, a surrogate model (SM) is utilized as the exact CAV model is not available. Performance dissimilarities between the SM and the specific CAV under evaluation, however, usually exist and can lead to the generation of suboptimal scenario library. The suboptimal library may increase the number of tests in order to reach a required evaluation precision, therefore may become the major source of evaluation inefficiency.

\begin{figure}[h!]
	\centering
	\includegraphics[width=0.48\textwidth]{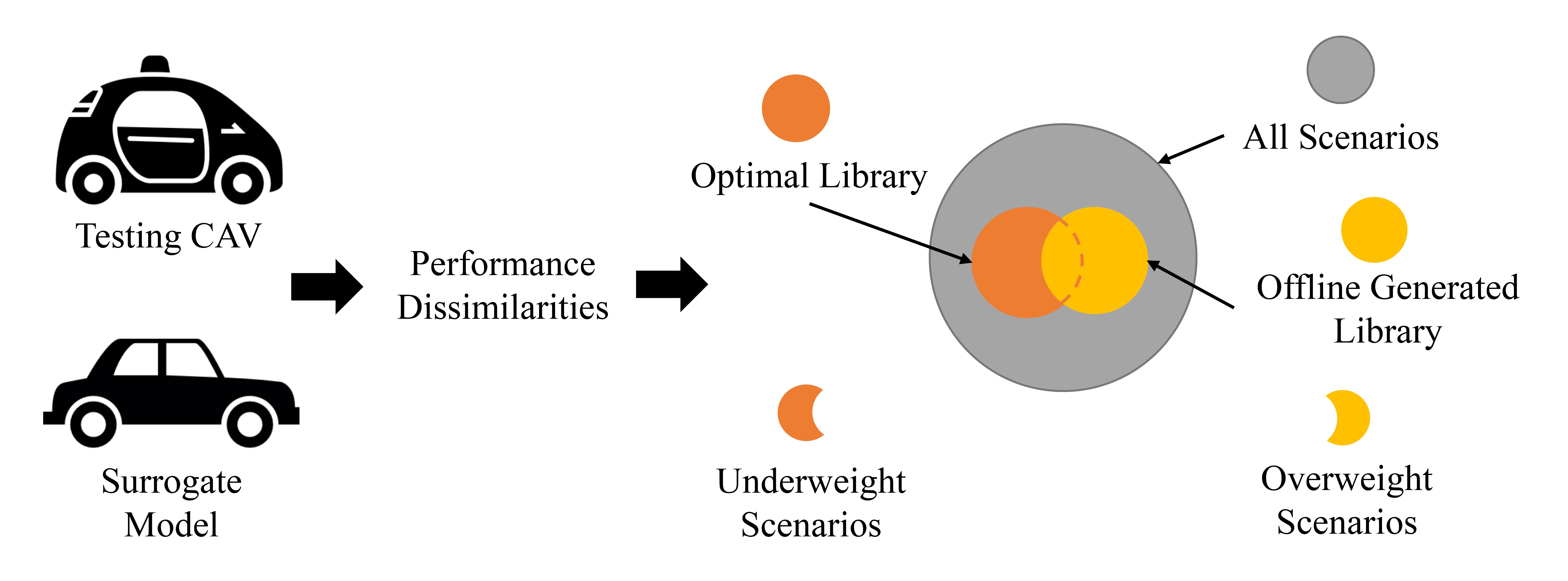}
	\caption{Illustration of suboptimal scenarios for a test CAV.}
	\label{fig_Dissimilarity}
\end{figure}

Two types of suboptimal scenarios exist, as shown in Fig. \ref{fig_Dissimilarity}. Underweight scenarios represent the critical scenarios that are ignored by the library, and overweight scenarios represent the uncritical scenarios that are included in the library. If we denote the scenario library generated by using the SM as ``offline generated library'', and a customized library that includes all critical scenarios specifically designed for a CAV as ``optimal library'',  the differences between these two libraries include both underweight and overweight scenarios.  

\begin{figure*}
	\centering
	\includegraphics[width=0.8\textwidth]{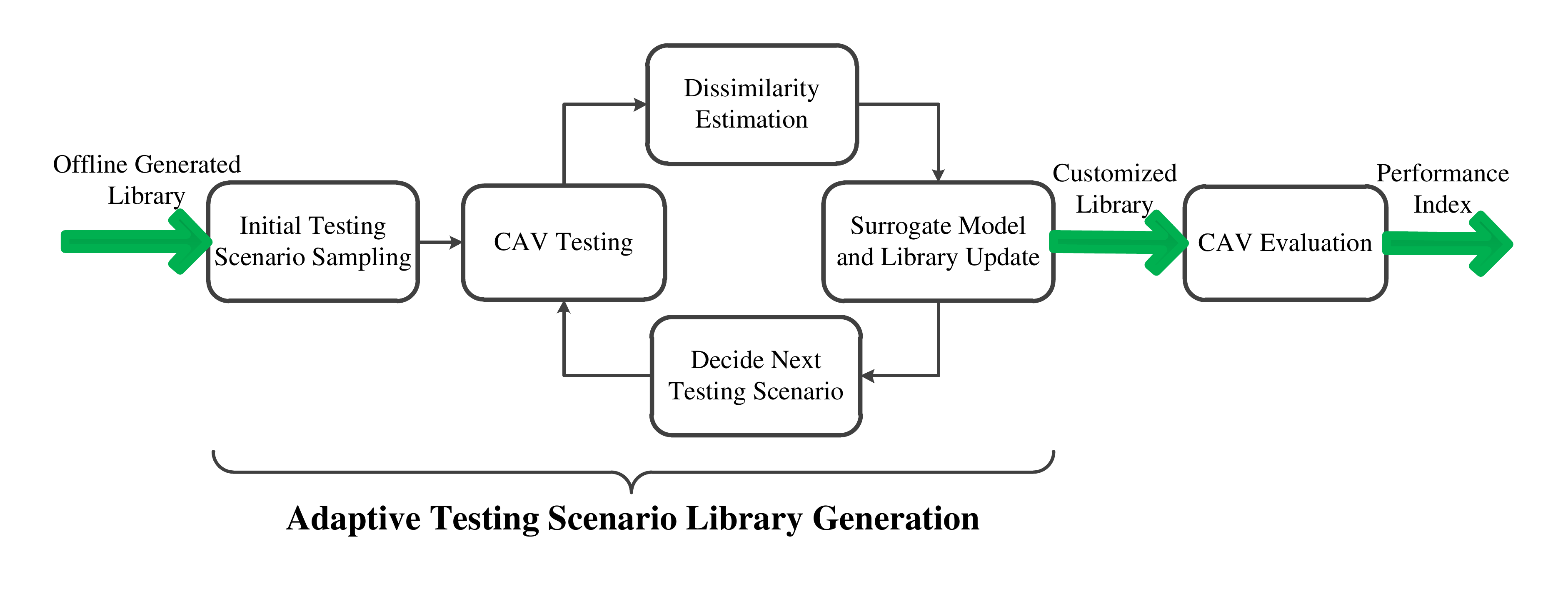}
	\caption{Illustration of the adaptive testing scenario library generation process.}
	\label{fig_Framework}
\end{figure*}

The goal of this paper is to generate the customized optimal library by reducing the number of suboptimal scenarios through an adaptive testing process. An illustration of this process is shown in Fig. \ref{fig_Framework}. The customization process starts with the test of CAV using a small set of scenarios sampled from the off-line generated library. After the initial testing, at each iteration, the most informative scenario is selected and tested, following that the SM is dynamically updated and the customized library is progressively improved, until the threshold for the dissimilarity compensation is reached. With the customized library, the CAV can be tested and evaluated in a more efficient manner, comparing with the evaluation method utilizing the offline generated library.

In the adaptive testing process, to leverage each CAV test, Bayesian optimization techniques \cite{snoek2012practical, frazier2018tutorial} are applied. The classification-based Gaussian Process Regression (GPR) \cite{williams2006gaussian} is used to estimate the nonstationary performance dissimilarities, and a new acquisition function is designed to determine the most informative testing scenario in each iteration. Both the prior knowledge (e.g., SM and offline generated library) and observations (e.g., results from the adaptive testing process) are utilized to customize the library. To validate the proposed framework, a cut-in case is studied in similar settings to those in \cite{feng2019testingII}. Comparing with the TSLG framework in \cite{feng2019testingI}, the new adaptive framework can further accelerate the evaluation process by a few orders of magnitude, e.g., 10-100.

The rest of this paper is organized as follows. For the convenience of the readers, Section \ref{sec_TSLG} briefly revisits the offline library generation method discussed in \cite{feng2019testingI, feng2019testingII, feng2020safety}. In Section \ref{sec_prob_form}, the problem of the adaptive testing process is formulated. The adaptive testing scenario library generation method is elaborated in Section \ref{sec_our_method}. In Section \ref{sec_case_study}, a cut-in case is presented to demonstrate the performance of the proposed method. Finally, Section \ref{sec_conclusion} concludes the paper.

\section{Revisit the TSLG Method}
\label{sec_TSLG}
The goal of the TSLG method \cite{feng2019testingI} is to generate a set of critical scenarios, which can be used to evaluate CAVs for certain performance indices more efficiently. If an event of interest is denoted as $A$, e.g., an accident event, the performance index can be defined as its occurrence probability:
\begin{eqnarray}
\label{eq_PA1}
P(A|\theta) = \sum_{x\in\B{X}} P(A|x,\theta)P(x|\theta),
\end{eqnarray}
where $x$ denotes the decision variables of the testing scenarios (e.g., maneuvers of background vehicles), $\B{X}$ denotes the feasible set of $x$, and $\theta$ denotes the pre-determined parameters by the ODD. Since $\theta$ keeps constant for a certain ODD, it will be omitted from now on to simplify the notations. So, the Eq. (\ref{eq_PA1}) is rewritten as
\begin{eqnarray}
\label{eq_PA2}
P(A) = \sum_{x\in\B{X}} P(A|x)P(x).
\end{eqnarray}

Essentially the on-road test is to evaluate the performance index in a naturalistic driving environment. Taking the cut-in case as an example, if a test CAV drives on public roads, experiences $n$ cut-in scenarios, and has $m$ accident events, the accident rate of the CAV in the cut-in scenarios is estimated as
\begin{eqnarray}
\label{eq_MCM}
P(A) &&= \sum_{x\in\B{X}} P(A|x)P(x), \nonumber\\
&&\approx \frac{1}{n} \sum_{i=1}^{n}P(A|x_i), x_i \sim P(x), \\
&&\approx \frac{m}{n}, \nonumber
\end{eqnarray}
where the last two equations are derived by Monte Carlo theory \cite{rubinstein2016simulation}. Here the cut-in scenarios on public roads follow the naturalistic distribution, i.e., $x_i \sim P(x)$. Because the accident event $A$ in the naturalistic driving environment is very rare, the required number of tests is intolerably large for reasonable estimation precision \cite{kalra2016driving}. We refer this as the rareness property in our paper.

To mitigate this issue, importance sampling techniques were applied by \cite{zhao2017accelerated} as
\begin{eqnarray}
\label{eq_IS}
P(A) &&= \sum_{x\in\B{X}} P(A|x)P(x), \nonumber\\
&&= \sum_{x\in\B{X}} \frac{P(A|x)P(x)}{q(x)}q(x), \\
&&\approx \frac{1}{n} \sum_{i=1}^{n}\frac{P(A|x_i)P(x_i)}{q(x_i)}, x_i \sim q(x), \nonumber
\end{eqnarray}
where $q(x)$ denotes an importance function satisfying
\begin{eqnarray}
q(x) \in [0,1], \sum_{x\in\B{X}} q(x) = 1, P(x)>0 \Rightarrow q(x)>0.
\end{eqnarray}
Comparing with Eq. (\ref{eq_MCM}), testing scenarios are sampled via the importance function $q(x)$ instead of $P(x)$. If $q(x)$ can increase the testing priority of critical scenarios, the evaluation efficiency can be improved. 

For a certain estimation precision, the minimal number of tests is determined by the importance function, and the required estimation precision can be measured by relative half-width for a given confidence level \cite{ross2017introductory}. With the confidence level at $100(1-\alpha)\%$, the relative half-width is defined as
\begin{eqnarray}
\label{eq_RHW}
l_r &&= \frac{\Phi^{-1}(1-\alpha/2)}{\mu_A} \sqrt{Var(\mu_A)}, \\
&&=\frac{\Phi^{-1}(1-\alpha/2)}{\mu_A} \frac{\sigma}{\sqrt{n}}, \nonumber
\end{eqnarray}
where $\mu_A=P(A)$, $\Phi^{-1}$ denotes the inverse cumulative distribution function of standard normal distribution $\C{N}(0,1)$, and $Var(\mu_A )=\sigma^2/n$ denotes the estimation variance. For a pre-determined half-width $\beta$, the minimal number of tests is derived as
\begin{eqnarray}
\label{eq_mTest}
n \ge \left(\frac{\Phi^{-1}(1-\alpha/2)}{\mu_A\beta}\right)^2 \sigma^2.
\end{eqnarray}
Therefore, the evaluation process has higher efficiency with a smaller $\sigma^2$. By the importance sampling theory \cite{owen2013monte}, the estimation variance can be derived as
\begin{eqnarray}
\label{eq_Var}
\sigma^2 = \sum_{x\in \B{X}} \frac{ \left(P(A|x)P(x)\right)^2 }{q(x)} - \mu^2_{A},
\end{eqnarray}
which is determined by the importance function. To obtain an importance function with small variance, a heuristic searching method was proposed in \cite{zhao2017accelerated}, which performs well in simple cases for safety evaluation. For complex cases and other metrics (e.g., functionality), the construction of a proper importance function remains a huge challenge.

To solve this problem, the scenario criticality was newly defined in \cite{feng2019testingI} as a combination of maneuver challenge ($P(S|x)$) and exposure frequency ($P(x)$) as 
\begin{eqnarray}
\label{eq_Value}
V(x) \overset{\rm def}{=} P(S|x) P(x), 
\end{eqnarray}
where $S$ denotes the event of interest with the SM of CAVs.  Integrated with a $\varepsilon$-greedy sampling policy, the importance function is essentially constructed as
\begin{eqnarray}
\label{eq_Prob_Sampling_2}
q(x) = \left\{
\begin{array}{ll}
(1-\epsilon)V(x)/W, &x \in \Phi\\
\epsilon / (N(\B{X}) - N(\Phi)), &x \notin \Phi
\end{array}
\right.
\end{eqnarray}
where $\Phi$ denotes the set of critical scenarios (i.e., the library), $N(\B{X})$ and $N(\Phi)$ denote the scenario numbers in the sets, and $W$ is a normalization factor as
\begin{eqnarray}
\label{eq_w}
W = \sum_{x \in \Phi} V(x).
\end{eqnarray}
The constructed importance function was justified by theoretical analysis and case studies regarding evaluation accuracy and efficiency in \cite{feng2019testingI, feng2019testingII, feng2020safety}. 

As discussed above, the maneuver challenge ($P(S|x)$) is evaluated by using an SM of CAV. However, performance dissimilarities between the SM and CAV models usually exist and can lead to the generation of suboptimal scenario library. The suboptimal library may increase the variance $\sigma^2$ and therefore decrease the evaluation efficiency. To further improve the evaluation efficiency, the problem of adaptive testing scenario library generation (ATSLG) is formulated and addressed in this paper.

\section{Problem Formulation}
\label{sec_prob_form}
In this section, the problem of ATSLG is formulated as a Bayesian optimization problem. Specifically, the ATSLG problem is analyzed in Subsection \ref{sec_ATSLG}. In Subsection \ref{sec_BOS}, the Bayesian optimization scheme is presented, and major challenges are analyzed.

\subsection{ATSLG Problem}
\label{sec_ATSLG}
The goal of the ATSLG is to minimize the estimation variance $\sigma^2$ by as few number of tests as possible. As discussed above, the key is to compensate for the performance dissimilarities between the SM and the CAV under test. The dissimilarity function is defined as
\begin{eqnarray}
\label{eq_dis_fun}
f(x) \overset{\rm def}{=} P(A|x) - P(S|x), x\in\B{X}.
\end{eqnarray}
Each test of the CAV will provide one observation of $f(x)$. 
Denote $\tilde{f}(x)$ as an estimation of $f(x)$, and then the SM can be updated with the compensation as
\begin{eqnarray}
\label{eq_SM_Up_1}
P(S'|x) = P(S|x) + \tilde{f}(x), x\in \B{X},
\end{eqnarray}
where $S'$ denotes the event of interest with the updated SM. 
The importance function can be constructed via the following equations:
\begin{eqnarray}
\label{eq_SM_IF}
\tilde{f}(x) \overset{\rm (\ref{eq_SM_Up_1})}{\to} P(S'|x) \overset{\rm (\ref{eq_Value})}{\to} V(x) \overset{\rm (\ref{eq_Prob_Sampling_2})}{\to} q(x),
\end{eqnarray}
and the estimation variance can be further obtained as 
\begin{eqnarray}
q(x) \overset{\rm (\ref{eq_Var})}{\to} \sigma^2.
\end{eqnarray}
Therefore, with the compensation $\tilde{f}$, the estimation variance should be reduced. If the mapping relation is denoted as a function $\sigma^2(\tilde{f})$, the ATSLG problem can be formulated as
\begin{eqnarray}
\label{eq_prob_ATSLG}
\min_{\tilde{f}\in \C{F}} \sigma^2(\tilde{f}),
\end{eqnarray}
where $\C{F}$ denotes the function space of $\tilde{f}$. 

As indicated in Theorem 2 in \cite{feng2019testingI}, the optimal solution of Eq. (\ref{eq_prob_ATSLG}) is obtained if the dissimilarities are exactly compensated, i.e., $\tilde{f}^*=f$. Generally, more observations of $f$ can lead to better compensation. However, each observation of $f$ required one real vehicle testing, which is time-consuming and cost-expensive. Therefore, the objective function should be optimized with as few observations as possible. 

To solve the problem, there are two critical subproblems. The first is how to select each test scenario $x$ for the new observation of $f(x)$. The informativeness of each scenario should be evaluated in the sense that how much information the new observation can provide for reducing the estimation variance. At each iteration, the most informative scenario should be selected for the next observation. The second is how to update the compensation function $\tilde{f}(x)$ for smaller $\sigma^2$ by leveraging all the existing observations and prior knowledge. 

\subsection{Bayesian Optimization Scheme}
\label{sec_BOS}
Bayesian optimization tries to optimize an unknown function $f(x)$ by as few observations as possible \cite{snoek2012practical}. It has been widely applied in various fields including intelligent transportation systems \cite{deshmukh2017testing, schultz2018bayesian, otsuka2019bayesian, chen2019bayesian, duan2020test, liu2020bayesian} (see \cite{frazier2018tutorial, shahriari2015taking} and references therein). It provides a powerful and flexible scheme especially for the optimization problems with expensive and black-box objective functions. The basic idea is to assume a prior probabilistic model for $f(x)$ and then exploit this model to decide where to observe $f(x)$ next, while integrating out uncertainty. Prior knowledge can be well utilized in the construction of the prior probabilistic model. To decide the next point for observation, various acquisition functions have been proposed for the measurement of the informativeness \cite{frazier2018tutorial}, e.g., expected improvement, knowledge gradient, entropy search, and predictive entropy search. With a properly designed acquisition function, the most informative scenario can be selected. Posterior knowledge can be obtained by integrating prior knowledge and observations.

In this paper, we propose to apply the Bayesian optimization scheme for the ATSLG problem. Specifically, the scheme of the ATSLG problem is described in Algorithm \ref{alg_BO_Scheme}. The SM and the offline generated library can be utilized as prior knowledge. The informativeness of each scenario can be evaluated by the acquisition function, and  $\tilde{f}(x)$ can be estimated as the posterior knowledge. Then, the SM as well as the library can be improved accordingly.

\begin{algorithm}[h!]
	\caption{Scheme of the ATSLG process.}
	\label{alg_BO_Scheme}
	\LinesNumbered 
	\KwIn{SM and offline generated library}
	\KwOut{Evaluation results of the CAV}
	{\bfseries Step 1}: Observe $f$ by testing the CAV with initial testing scenarios. ({\emph{Sec \ref{sec_initial_sampling}}}) \\
	{\bfseries Step 2}: 
	\While{stop criterion is not satisfied}{
		{\bfseries Step 2.1}: Obtain the estimation $\tilde{f}$ ({\emph{Sec \ref{sec_dissi_est}}})\;
		{\bfseries Step 2.2}: Update SM and library ({\emph{Sec \ref{sec_SM_up}}})\;
		{\bfseries Step 2.3}: Decide next iteration of testing scenarios ({\emph{Sec \ref{sec_acquisition}}})\;
		{\bfseries Step 2.4}: Observe $f$ by testing the CAV with new scenarios\;
	}
	{\bfseries Step 3}: Test and evaluate the CAV with the customized library ({\emph{Sec \ref{sec_algorithm}}}).
\end{algorithm}

When applying the Bayesian optimization scheme to the ATSLG problem, there are three major challenges as follows:

First, the ATSLG problem optimizes in the function space, $\tilde{f}\in\C{F}$, instead of the parameter space, $x \in \B{X}$, as shown in Eq. (\ref{eq_prob_ATSLG}). \textcolor{black}{Essentially, the function space is infinite-dimensional, and optimization in the function space belongs to the domain of infinite dimensional analysis \cite{guide2006infinite}. For the common Bayesian optimization problems, however, the decision variable $x \in \B{X}$ is finite-dimensional, which is less complex and challenging. Although the function space can be simplified as a finite-dimensional space after the discretization, its dimension is still much higher than the decision variable. In the cut-in case of this paper, for example, the dimension of $\tilde{f}$ is 3,420 after discretization, while the dimension of $x$ is only 2.}

Second, performances of a CAV may change more drastically in certain scenario neighborhoods than others, and therefore the covariance of the dissimilarity function can be highly non-stationary and nonlinear. 

Third, the objective function $\sigma^2$ is unavailable for the ATSLG problem. As shown in Eq. (\ref{eq_Var}), $\sigma^2$ cannot be calculated unless $\mu_A$ is known, which is exactly what needs to be evaluated. However, most existing acquisition functions of Bayesian optimization methods are calculated based on the availability of objective functions. Consequently, a new acquisition function needs to be designed.

We aim to address the above challenges in the following section.

\section{Adaptive Testing Scenario Library Generation}
\label{sec_our_method}
In Subsection \ref{sec_initial_sampling}, to ``prime the pump'' with initial testing scenarios, a sampling mechanism that balances the exploitation of the offline generated library and exploration outside the library is designed. Such a sampling mechanism will provide a sketch of the dissimilarity function. In Subsection \ref{sec_dissi_est}, different from most Bayesian optimization methods where explicit objective functions are estimated, the dissimilarity function is estimated by the Gaussian process regression (GPR) method. To handle the non-stationary challenge, scenarios are classified into two groups before applying the GPR method, resulting in the classification-based GPR method. In Subsection \ref{sec_SM_up}, the SM is compensated with the estimated dissimilarity function, and the new library is generated accordingly. Furthermore, in Subsection \ref{sec_acquisition}, the informativeness of each scenario is measured by the estimated improvement of the evaluation efficiency, and then a new acquisition function is designed. Finally, the overall algorithm is summarized in Subsection \ref{sec_algorithm}.

\subsection{Initial Testing Scenarios} 
\label{sec_initial_sampling}
To provide a sketch of the dissimilarity function, we should balance the exploitation of the offline generated library and exploration outside the library. To this end, a simple yet effective policy is proposed as follows. Since scenarios of the library have higher testing priority, they are more likely to be overweighted. To find overweight scenarios, the library is sampled according to scenario criticality values. Similarly, scenarios outside the library are more likely to be underweighted. To find underweight scenarios, scenarios outside the library are randomly sampled with a probability $\gamma$. Comparing with the $\epsilon$ in Eq. (\ref{eq_Prob_Sampling_2}), the value of $\gamma$ is much larger, e.g., 0.5. Similar to the ``No Free Lunch Theorem'' \cite{wolpert1997no}, if there is no additional information about locations of the underweight scenarios, any searching scheme is no better than random sampling. Incorporating all these considerations, the initial testing scenarios are sampled as
\begin{eqnarray}
P(x_0) = \left\{
\begin{array}{ll}
(1-\gamma)V(x_0)/W, &x_0 \in \Phi,\\
\gamma / (N(\B{X}) - N(\Phi)), &x_0 \in \B{X} \backslash \Phi,
\end{array}
\right.
\end{eqnarray}
where $x_0$ denotes an initial testing scenario. 

\subsection{Dissimilarity Function Estimation}
\label{sec_dissi_est}
The dissimilarity function is estimated by the GPR method \cite{williams2006gaussian}, because of the following advantages. As a non-parametric method, it is not limited by a functional form and thus is flexible and powerful for estimating highly nonlinear functions. Moreover, it is also convenient to add prior knowledge of the specific problem by selecting different covariance functions. In this paper, a non-stationary covariance function is designed by the classification-based GPR method. Furthermore, besides the function estimation, it can also provide a probability distribution over the function estimation, which captures the estimation uncertainty. The informativeness of each scenario can be evaluated based on the estimation uncertainty. 

The basic idea is to use a Gaussian process (GP) to describe a probability distribution over the functions. Specifically, the value of $f(x)$ at each scenario $x$ is viewed as a Gaussian random variable, and values of $f(x)$ at all scenarios follow a joint Gaussian distribution. As a result, $f(x)$ can be represented by the GP as
\begin{eqnarray}
\label{eq_GP}
f(x) \sim \C{GP}\left(  m(x), k(x, x') \right),
\end{eqnarray}
where both $x$ and $x'$ denote scenarios, $m(x)$ denotes the mean function, and $k(x,x')$ denotes the covariance function.

Based on the GP, the values of $f(x)$ for unobserved scenarios can be estimated by the regression, namely the GPR. Denote $N$ points of scenarios with observations as $\B{X}_N = \{ x_n \in  \B{X} \}^N_{n=1}$, and $N^*$ points of scenarios without observations as $\B{X}_{N^*} = \{ x_{n^*} \in  \B{X} \}^{N^*}_{n^*=1}$. An observation of $f(x)$ is equivalent to one test of the CAV, and the observation results are denoted as $f(\B{X}_N )$. As elaborated in \cite{williams2006gaussian}, $f(\B{X}_{N^*})$ can be estimated by the posterior probability distribution as
\begin{eqnarray}
\label{eq_GPR}
f(\B{X}_{N^*})|f(\B{X}_N ) \sim \C{GP} \left( \tilde{f}_{\B{X}_N}(\B{X}_{N^*}), \sigma^2_{P, \B{X}_N}(\B{X}_{N^*}) \right),
\end{eqnarray}
where the mean $\tilde{f}_{\B{X}_N}(\B{X}_{N^*})$ indicates the estimation, and the variance $\sigma^2_{P, \B{X}_N}(\B{X}_{N^*})$ indicates the estimation uncertainty.

A nonstationary covariance function is designed by incorporating the Gaussian process classification (GPC). \textcolor{black}{As discussed in \cite{shahriari2015taking}, there are various approaches for designing a nonstationary covariance function, such as nonstationary kernels, partitioning, and heteroscedasticity. Here we utilize the idea of partitioning,} which divides the variable space into several regions and applied GPR in each region respectively to handle the nonstationary issue. Different from the deterministic classification method, GPC provides a probability distribution of different classes for each variable. As a result, a variable could belong to multiple classes with different probabilities and, therefore, be estimated by the GPR in each class respectively. The final estimation of the variable is the expectation of all these estimation results. In this paper, scenarios are divided into two classes,  suboptimal scenarios ($y(x)=+1$) and optimal scenarios ($y(x)=-1$), by the values of $f(x)$ as 
\begin{eqnarray}
\label{eq_class}
y(x) = \left\{\begin{matrix}
+1, &f(x)\neq0\\
-1, &f(x)=0
\end{matrix}
\right..
\end{eqnarray}
The class labels of the scenarios $\B{X}_N$, i.e., $y(\B{X}_N)$, are calculated based on the observations. Let $\B{X}_{N_1}$ denote the observed suboptimal scenarios  and $\B{X}_{N_2}$ denote the observed optimal scenarios. To classify the unobserved scenarios, $y(\B{X}_{N^*})$ can be estimated by the posterior probability as
\begin{eqnarray}
\label{eq_P_C}
P\left( y(x)=+1 | y(\B{X}_N) \right), x \in \B{X}_{N^*},
\end{eqnarray}
where the analytic equations can be found in \cite{williams2006gaussian}. For notation simplification, Eq. (\ref{eq_P_C}) is denoted as $P_{1, \B{X}_N}(x)$, and 
 \begin{eqnarray}
 P_{2, \B{X}_N}(x) = 1- P_{1, \B{X}_N}(x).
 \end{eqnarray}

Finally, the GPC-based GPR results of $f(x)$ can be represented as
\begin{eqnarray}
\label{eq_GPR_fx}
&&f_{\B{X}_N}(x) \sim \\
&&\ \ \ \ \ \ \left\{ \begin{matrix}
\C{N}\left(\tilde{f}_{\B{X}_{N_1}}(x), \sigma^2_{P, \B{X}_{N_1}}(x)\right), &\text{with } P_{1, \B{X}_N}(x),\\
\C{N}\left(\tilde{f}_{\B{X}_{N_2}}(x), \sigma^2_{P, \B{X}_{N_2}}(x)\right), &\text{with } P_{2, \B{X}_N}(x),
\end{matrix}  \right. \nonumber
\end{eqnarray}
where $\C{N}(\tilde{f}_{\B{X}_{N_1}}(x), \sigma^2_{P, \B{X}_{N_1}}(x))$ denotes the GPR results in suboptimal scenarios, and $\C{N}(\tilde{f}_{\B{X}_{N_2}}(x), \sigma^2_{P, \B{X}_{N_2}}(x))$ denotes the results in optimal scenarios. The estimation of $f(x)$ can be obtained by the expectation as
\begin{eqnarray}
\label{eq_GPR_fx_E}
\tilde{f}_{\B{X}_{N}}(x) = P_{1, \B{X}_N}(x) \tilde{f}_{\B{X}_{N_1}}(x) + P_{2, \B{X}_N}(x) \tilde{f}_{\B{X}_{N_2}}(x).
\end{eqnarray}

\subsection{Surrogate Model Update and Library Generation}
\label{sec_SM_up}
One limitation of the GPR method is the Gaussian assumption, which would produce a huge number of small yet non-zero values. It is inconsistent with the rareness property of the SM that most values are zero. To maintain the rareness property, a scenario is set as uncritical, if both prior and posterior knowledge indicate it is very likely to be uncritical. 

Specifically, with the compensation $\tilde{f}_{\B{X}_{N}}(x)$ in Eq. (\ref{eq_GPR_fx_E}), the SM is updated by
\begin{eqnarray}
\label{eq_SM_E}
P_E(S_{\B{X}_N}|x) = \left\{ \begin{matrix}
P(S|x) + \tilde{f}_{\B{X}_{N}}(x), &x \in \B{X}/\B{U},\\
0, &x \in \B{U},
\end{matrix}  \right.
\end{eqnarray}
where the set $\B{U}$ is defined to keep the rareness property. It is defined as
\begin{eqnarray}
\label{set_U}
\B{U} = \left\{ x \in \B{X}: P(S|x)=0, P_{1, \B{X}_N}(x) \le P_{th} \right\},
\end{eqnarray}
where $P_{th}$ is a pre-determined probability threshold for classification, e.g., 0.5. Scenarios $x\in\B{U}$ are indicated uncritical by both the prior knowledge ($P(S|x)=0$) and the posterior knowledge ($P_{1, \B{X}_N}(x)\le P_{th}$). Based on the updated SM, a new importance function $q_{\B{X}_N}(x)$, as well as a library, can be constructed by Eq. (\ref{eq_SM_IF}).

\subsection{Acquisition Function Design}
\label{sec_acquisition}
The acquisition function should be designed to measure the informativeness of each scenario for selecting the next test scenario. Since the objective function, $\sigma^2$, is unavailable, a surrogate measure is designed by the estimated reduction of $\sigma^2$. Based on the surrogate measure, a new acquisition function is designed leveraging the dissimilarity function estimation and the estimation uncertainty.

As indicated in Theorem 2 in \cite{feng2019testingI}, Eq. (\ref{eq_Var}) can be approximated by 
\begin{eqnarray}
\sigma^2 \approx \sum_{x\in\B{X}} \frac{P^2(x)}{q(x)}  f^2(x).
\end{eqnarray}
The reduction of $\sigma^2$ for each testing scenario can be approximated by the surrogate measure $\frac{P^2(x)}{q(x)} f^2(x)$. Based on the estimation results $f_{\B{X}_N}(x)$ in Eq. (\ref{eq_GPR_fx}), the informativeness of each scenario can be evaluated by its expectation over the classification probability and the estimation uncertainty as 
\begin{eqnarray}
\label{eq_EPI}
EI_{\B{X}_N}(x)  \overset{\rm def}{=} E\left[ \frac{P^2(x)}{q_{\B{X}_N}(x)}f^2_{\B{X}_N}(x) \right],
\end{eqnarray}
where $q_{\B{X}_N}(x)$ denotes the updated importance function according to Eq. (\ref{eq_SM_IF}). Applying the integration by parts and Eq. (\ref{eq_GPR_fx}), the analytical form of Eq. (\ref{eq_EPI}) can be derived as 
\begin{eqnarray}
EI_{\B{X}_N}(x) = \frac{P^2(x)}{q_{\B{X}_N}(x)} \left[ P_{1, \B{X}_N}(x) E_{1} + P_{2, \B{X}_N}(x) E_{2} \right],
\end{eqnarray}
where
\begin{eqnarray}
E_{i}  \overset{\rm def}{=} \tilde{f}^2_{\B{X}_{N_i}}(x) +  \sigma^2_{P, \B{X}_{N_i}}(x).
\end{eqnarray}

To better explore the boundaries of the classification, the classification variance $\sigma^2_{C, \B{X}_N}(x)$ is further incorporated as
\begin{eqnarray}
I_{\B{X}_N}(x) = w \frac{EI_{\B{X}_N}(x)}{U_E} + \frac{\sigma^2_{C, \B{X}_N}(x)}{U_C}, 
\end{eqnarray}
where $w$ is a weight to balance the two terms, and $U_E$, $U_C$ are normalization factors to make the metrics comparable. The classification variance can be calculated by the GPC method \cite{gramacy2005bayesian}. Recall that the scenarios $x \in \B{U}$ are indicated uncritical. Therefore, the acquisition function, which exploits existing information, is unlikely to explore these scenarios. To search possible ``unexpected'' suboptimal scenarios, a small probability ($\beta$) of random sampling is applied. Finally, the next iteration of testing scenario is decided by
\begin{eqnarray}
\label{eq_next_sce}
x_{N+1}=\left\{ \begin{matrix}
\max_{x} I_{\B{X}_N}(x), x\in \B{X}/\B{U}, & \text{with }  1-\beta\\
\text{random sampling for }x\in \B{U}, &\text{with } \beta 
\end{matrix}\right..
\end{eqnarray}

\subsection{Overall Algorithm}
\label{sec_algorithm}
As shown in Fig. \ref{fig_Framework} and Algorithm 1, the test of a CAV includes three steps, described in the following:

The first step is to test the CAV with initial scenarios generated as in Subsection \ref{sec_initial_sampling}. The testing results provide a sketch of the dissimilarity function. 

Based on the sketch, the second step is to test the CAV with the most informative scenario iteratively. At each iteration, the dissimilarity function is estimated as in Subsection \ref{sec_dissi_est}, the SM as well as the library is updated as in Subsection \ref{sec_SM_up}, and the acquisition function is calculated to determine the next test scenario as in Subsection \ref{sec_acquisition}. The iterative process will stop if the number of tests is larger than the pre-determined budget or the estimation precision is satisfied. 

With the updated library, the third step is to test and evaluate the CAV with the $\epsilon$-greedy sampling policy as shown in Eq. (\ref{eq_Prob_Sampling_2}). The minimal number of tests can be determined by Eq. (\ref{eq_mTest}), and the CAV performance can be evaluated by Eq. (\ref{eq_IS}).

\section{Cut-in Case Study}
\label{sec_case_study}
In this section, the proposed method is demonstrated in a cut-in case for safety evaluation.  

\subsection{Case Description}
\begin{figure}[h!]
	\centering
	\includegraphics[width=0.2\textwidth]{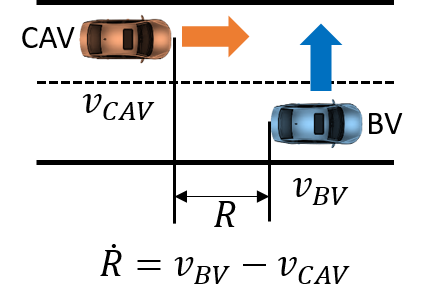}
	\caption{Illustrations of the cut-in case.}
	\label{fig_Case}
\end{figure}

The cut-in case is illustrated in Fig. \ref{fig_Case} (a), where a background vehicle (BV) makes a lane change in front of the test CAV. Similar to previous work \cite{zhao2017accelerated, feng2019testingII}, the decision variables are constructed as 
\begin{eqnarray}
x = (R, \dot{R}),
\end{eqnarray}
where $R$ and $\dot{R}$ denote the range and range rate (the longitudinal speed difference) of the two vehicles at the cut-in moment. The accident event is defined as the minimal distance between the two vehicles is smaller than a threshold, i.e., $d_{min}=1m$. The safety performance is evaluated by the accident rate of the CAV on public roads. A CAV car-following model used in \cite{zhao2017accelerated, feng2019testingII}, which combines adaptive cruise control and autonomous emergency braking functions, is evaluated.

\subsection{Offline Library Generation}
\begin{figure}[h!]
	\centering
	\begin{minipage}{.49\linewidth}
		\centering
		\includegraphics[width=1\textwidth]{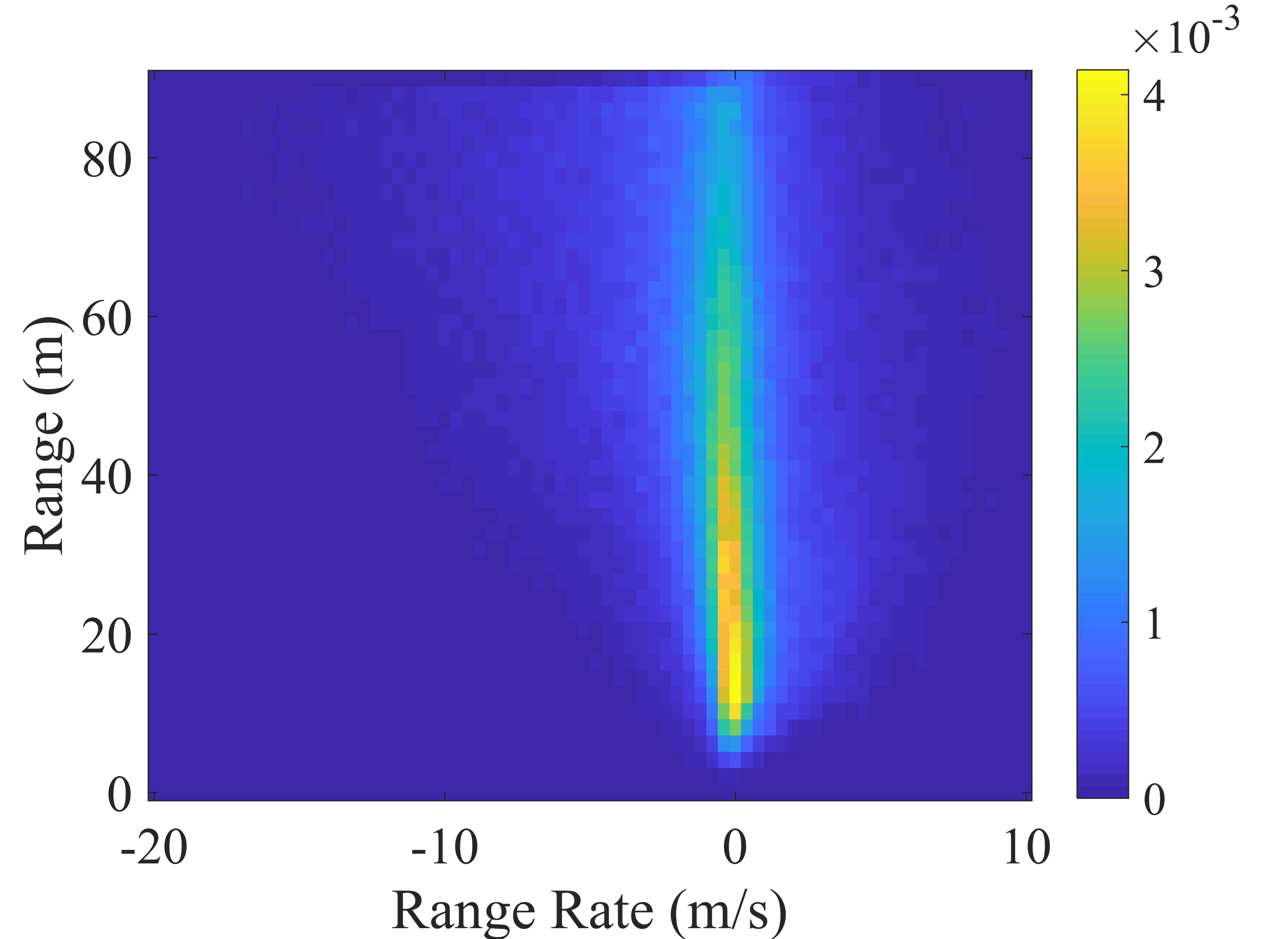}
		\centerline{\small (a) Exposure Frequency}
	\end{minipage}
	\begin{minipage}{.49\linewidth}
		\centering
		\includegraphics[width=1\textwidth]{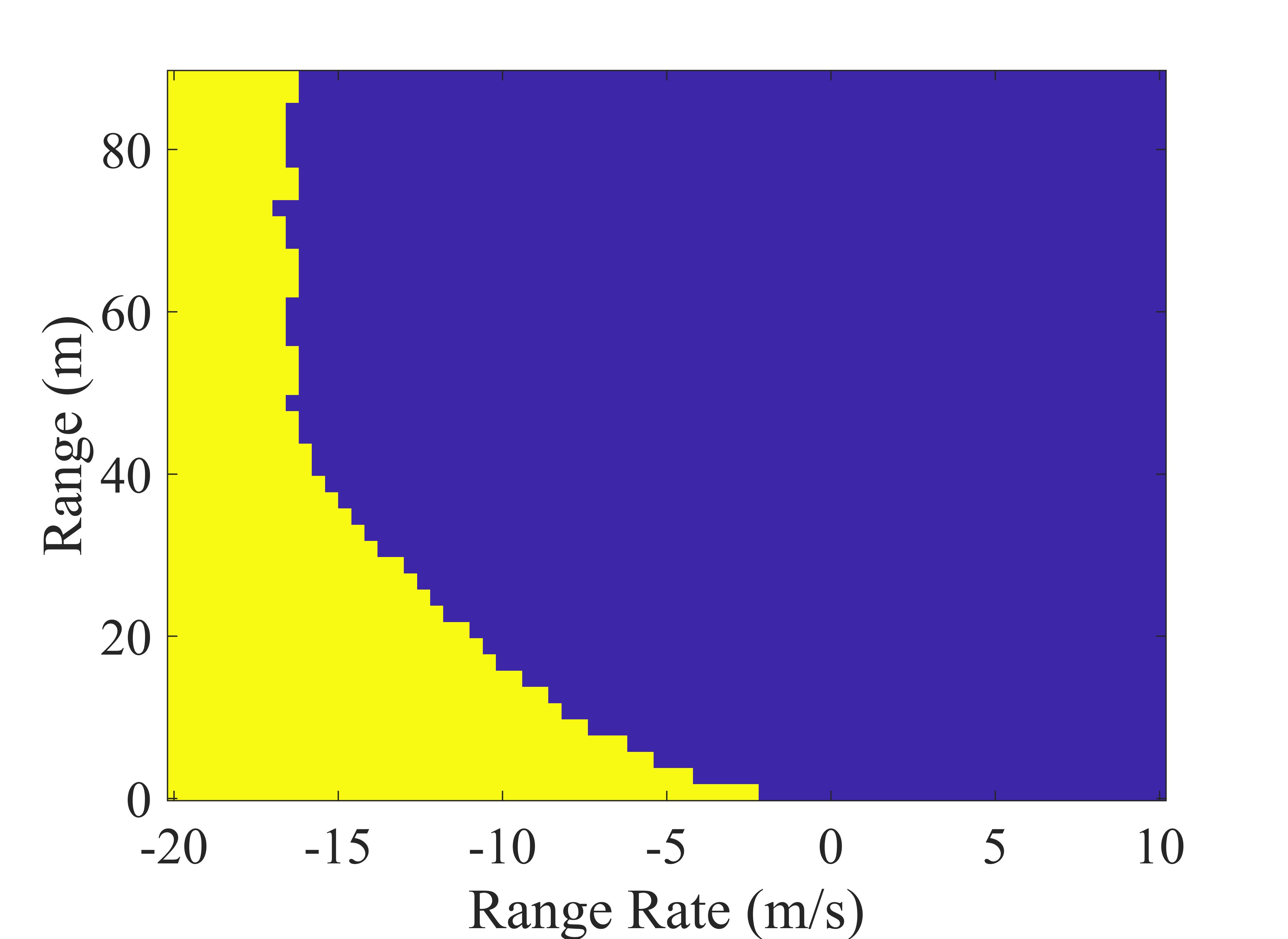}
		\centerline{\small (b) Maneuver Challenge}
	\end{minipage}
	\caption{The exposure frequency and maneuver challenge of the cut-in case based on the FVDM.} 
	\label{fig_Px}
\end{figure}

The TSLG method in \cite{feng2019testingI} is conducted to generate the offline library.  To estimate the exposure frequency of the cut-in scenarios, NDD from the Safety Pilot Model Deployment program at the University of Michigan \cite{bezzina2014safety} is utilized. A total number of 414,770 qualified cut-in events are successfully obtained. The joint probability distribution of the cut-in range and range rate (i.e., P(x)) is shown in Fig. \ref{fig_Px} (a). 

To determine the maneuver challenge, the Full Velocity Difference Model (FVDM) is adopted as the SM because it is one of the most widely used car-following models representing human drivers \cite{ro2017formal}. It is worth noting that, to make the dissimilarity prominent, the selected SM in this case is different from the Intelligent Driving Model adopted in \cite{feng2019testingII}. Specifically, the car-following acceleration is determined by
\begin{eqnarray}
u(k+1) = C_0 \left[  V_1 + V_2 \tanh( C_1(R(k)-L)-C_2 )-\dot{R}(k) \right], \nonumber
\end{eqnarray}
where $u(k+1)$ denotes the acceleration of the CAV at time step $k+1$, $C_0$, $V_1$, $V_2$, $C_1$, $L$, and $C_2$ are constant parameters. Similar to \cite{hamdar2008existing}, the constraints of acceleration and velocity are added to make the model more practical, i.e., model accident-prone behaviors, as
\begin{eqnarray}
v_{min} \le v \le v_{max}, a_{min} \le u \le a_{max}.
\end{eqnarray}
All calibrated parameters in \cite{ro2017formal} are adopted as listed in Table \ref{tab_cutin}. Fig. \ref{fig_Px} (b) shows the safety performance of the constructed SM, where the SM has accidents in the yellow region. 

\linespread{1.2}
\begin{table}[h!]
	\centering
	\footnotesize
	\setlength{\abovecaptionskip}{1pt}
	\setlength{\belowcaptionskip}{3pt}	
	\caption{The parameter values of the cut-in case. }
	\label{tab_cutin}
	\begin{tabular}{cccc}
		\hline
		\multicolumn{1}{c}{\bfseries Parameter } &   \multicolumn{1}{c}{ \bfseries Value} &\multicolumn{1}{c}{\bfseries Parameter } &   \multicolumn{1}{c}{ \bfseries Value}  \\ \hline
		$C_0$  &0.85 &$V_1$ &6.75 \\ 
		$V_2$ &7.91 &$C_1$ &0.13\\ 	
		$L$ &5 &$C_2$ &1.57\\ 
		$v_{min}$ &2 &$v_{max}$ &40\\
		$a_{max}$ &2&$a_{min}$ &-4\\ 
		$P_{th}$ &0.7 & $w$ &0.5\\
		$\gamma$ &0.5 &$\epsilon$ &0.1\\	 \hline
	\end{tabular}
\end{table}
\linespread{1.0}

\begin{figure}[h!]
	\centering
	\includegraphics[width=0.35\textwidth]{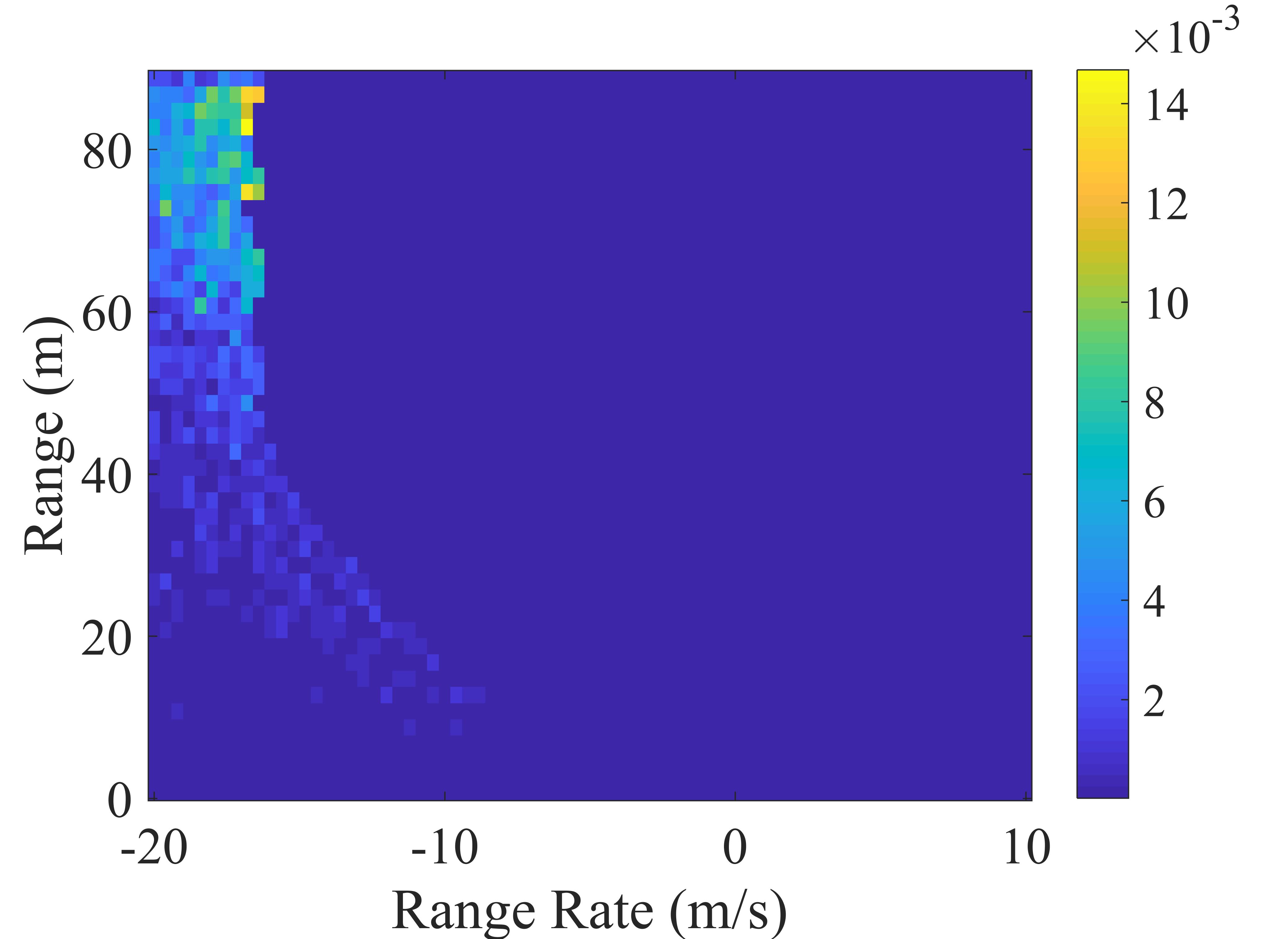}
	\caption{The offline generated library of the cut-in case for safety evaluation based on the FVDM.} 
	\label{fig_OffLib}
\end{figure}

To obtain critical scenarios and construct the library, the threshold for critical scenarios is determined as
\begin{eqnarray}
V(x) > \frac{1}{N(\B{X})} = 2.9 \times 10^{-4},
\end{eqnarray}
where $N(\B{X})$ denotes the total number of scenarios, and $N(\B{X}) =47\times76=3,420$. The range and range rate are discretized by $2m$ and $0.4m/s$ respectively, and their boundaries are $(0,90]$ and $[-20,10]$. Fig. \ref{fig_OffLib} shows the obtained probability distribution combining both exposure frequency and maneuver challenge. The colors denote the sampling probabilities of the scenarios. In this case, the generated library contains a total number of 342 critical scenarios, which is about 10\% of all scenarios.

\subsection{Adaptive Library Generation}
After the offline scenario library is generated, 50 scenarios are sampled as initial testing scenarios, and then 50 iterations of adaptive testing are conducted. \textcolor{black}{As discussed in Section \ref{sec_algorithm}, at each iteration, the CAV is tested for one time in the selected scenario.} The MATLAB toolbox in \cite{rasmussen2003gaussian} is utilized to execute the GPR and GPC. The squared exponential with automatic relevance determination covariance function is applied for the regression and classification as
\begin{eqnarray}
k(x,x') = \sigma^2_f \exp \left[ -\frac{1}{2} \sum_{d=1}^{D} \left(\frac{x_d - x'_d}{\lambda_d}\right)^2 \right],
\end{eqnarray}
where $D$ denotes the dimensions of $x$. $\sigma_f$ and  $\lambda_d$ are hyper-parameters. As pointed out in \cite{williams2006gaussian}, the squared exponential function is probably the most widely used covariance function, and the automatic relevance determination is usually used for determining the hyper-parameters of the specific problem. Please note that this covariance function is neither unique nor optimal for the problem, and further investigation is required for the design of better covariance functions. The computation is conducted with MATLAB 2017a, in a workstation equipped with Intel i7-7700 CPU and 16G RAM, and takes about 48 seconds in total.

\begin{figure}[h!]
	\centering
	\begin{minipage}{.49\linewidth}
		\centering
		\includegraphics[width=0.98\textwidth]{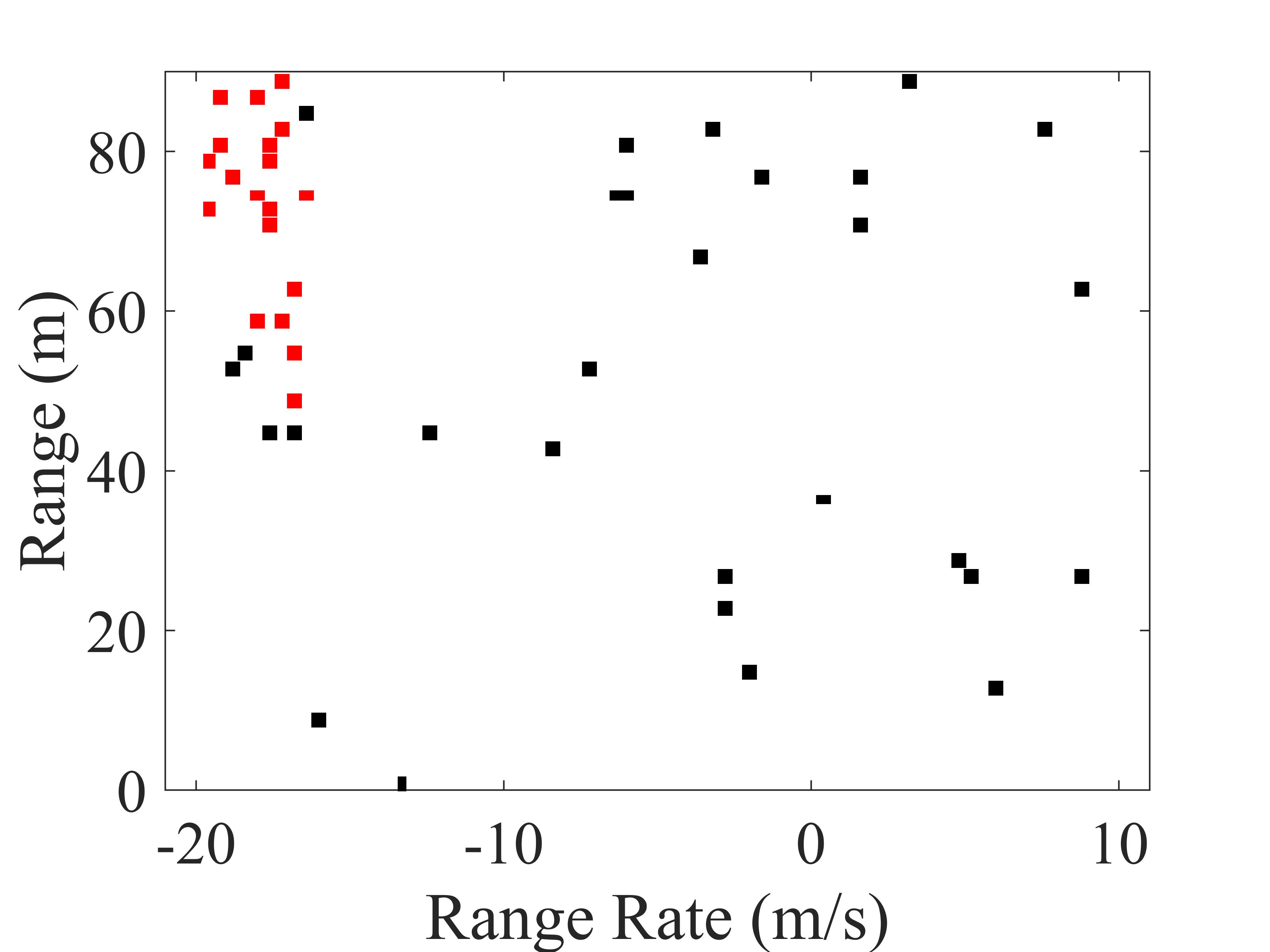}
		\centerline{\small (a) Initial Testing Results}
	\end{minipage}
	\begin{minipage}{.49\linewidth}
		\centering
		\includegraphics[width=1\textwidth]{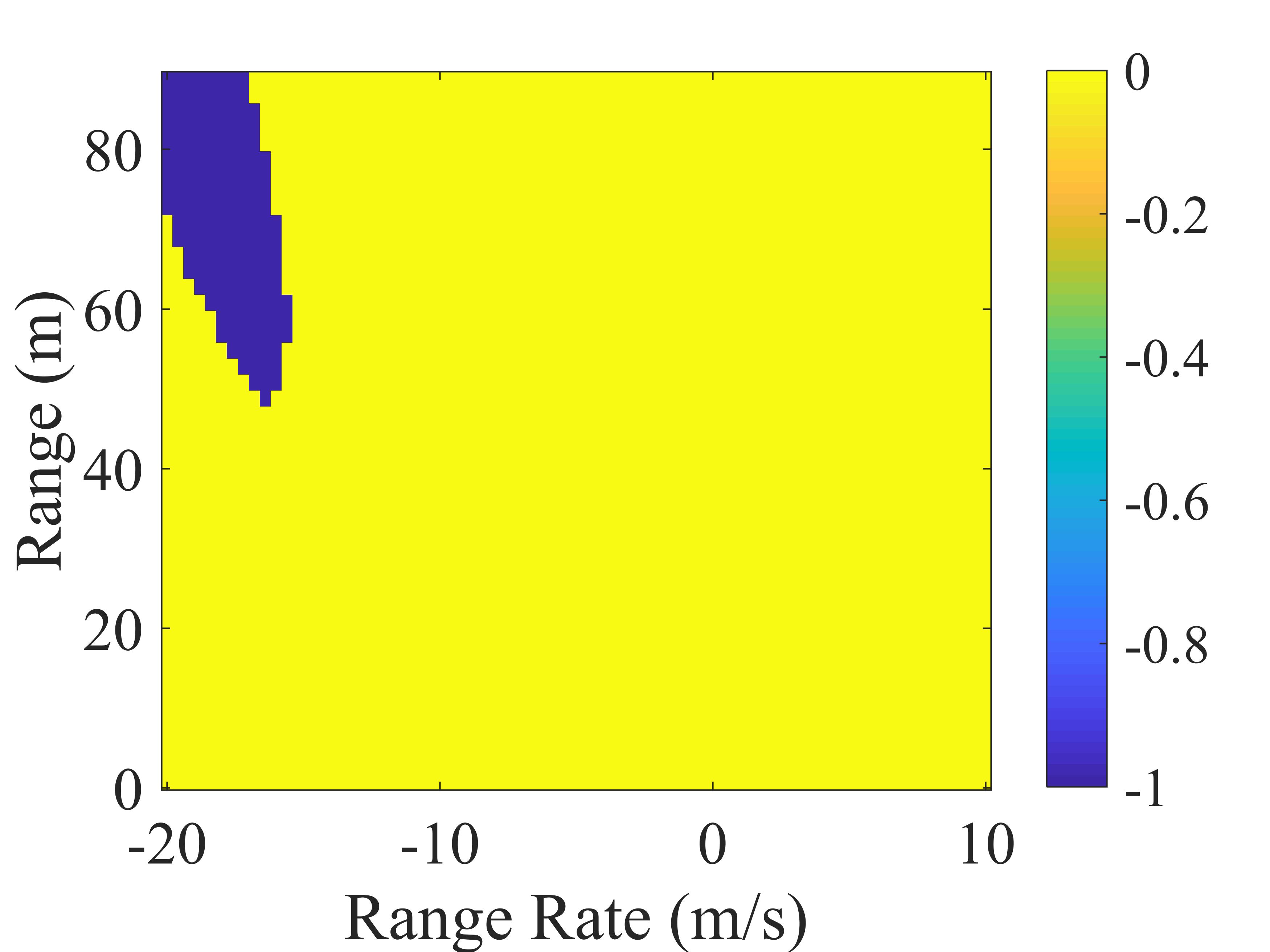}
		\centerline{\small (b) Proposed Method}
	\end{minipage}
	\begin{minipage}{.49\linewidth}
		\centering
		\includegraphics[width=1\textwidth]{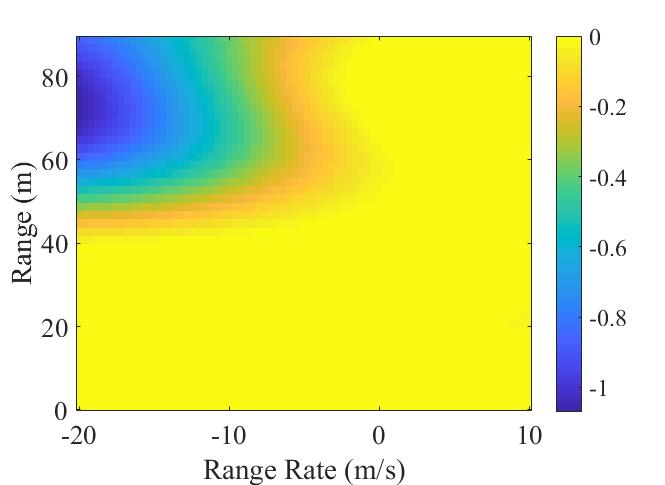}
		\centerline{\small (c) GPR Method \cite{rasmussen2003gaussian}}
	\end{minipage}
	\begin{minipage}{.49\linewidth}
		\centering
		\includegraphics[width=1\textwidth]{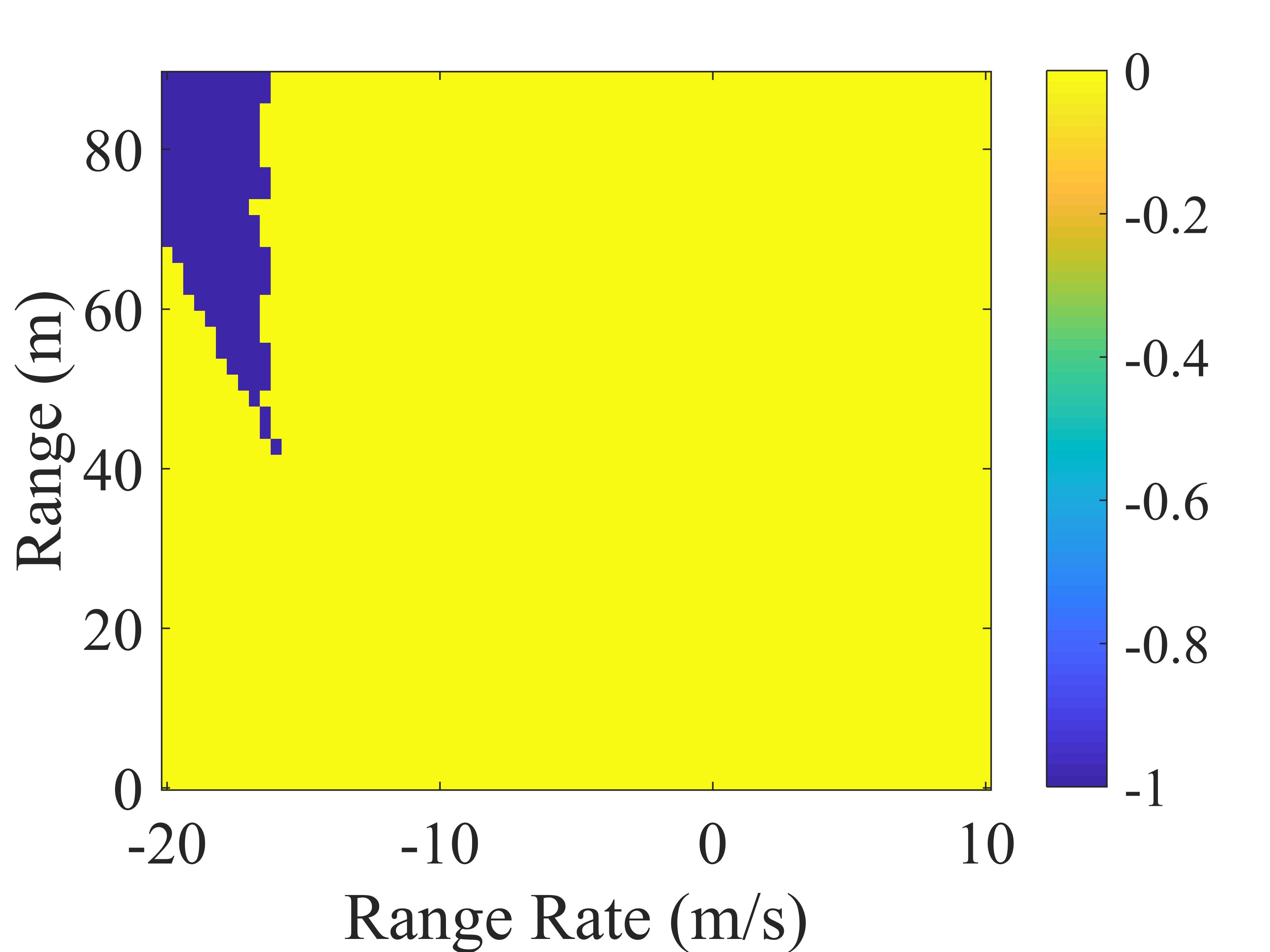}
		\centerline{\small (d) Ground Truth}
	\end{minipage}
	\caption{\color{black}(a) Testing results of the initial testing scenarios including the observed suboptimal scenarios (red dots) and the observed optimal scenarios (black dots); (b) Regression results by the proposed method; (c) Regression results by the GPR method \cite{rasmussen2003gaussian};  (d) Ground truth of the dissimilarity function.} 
	\label{fig_IniSam}
\end{figure}

\begin{figure}[h!]
	\centering
	\begin{minipage}{.49\linewidth}
		\includegraphics[width=1\textwidth]{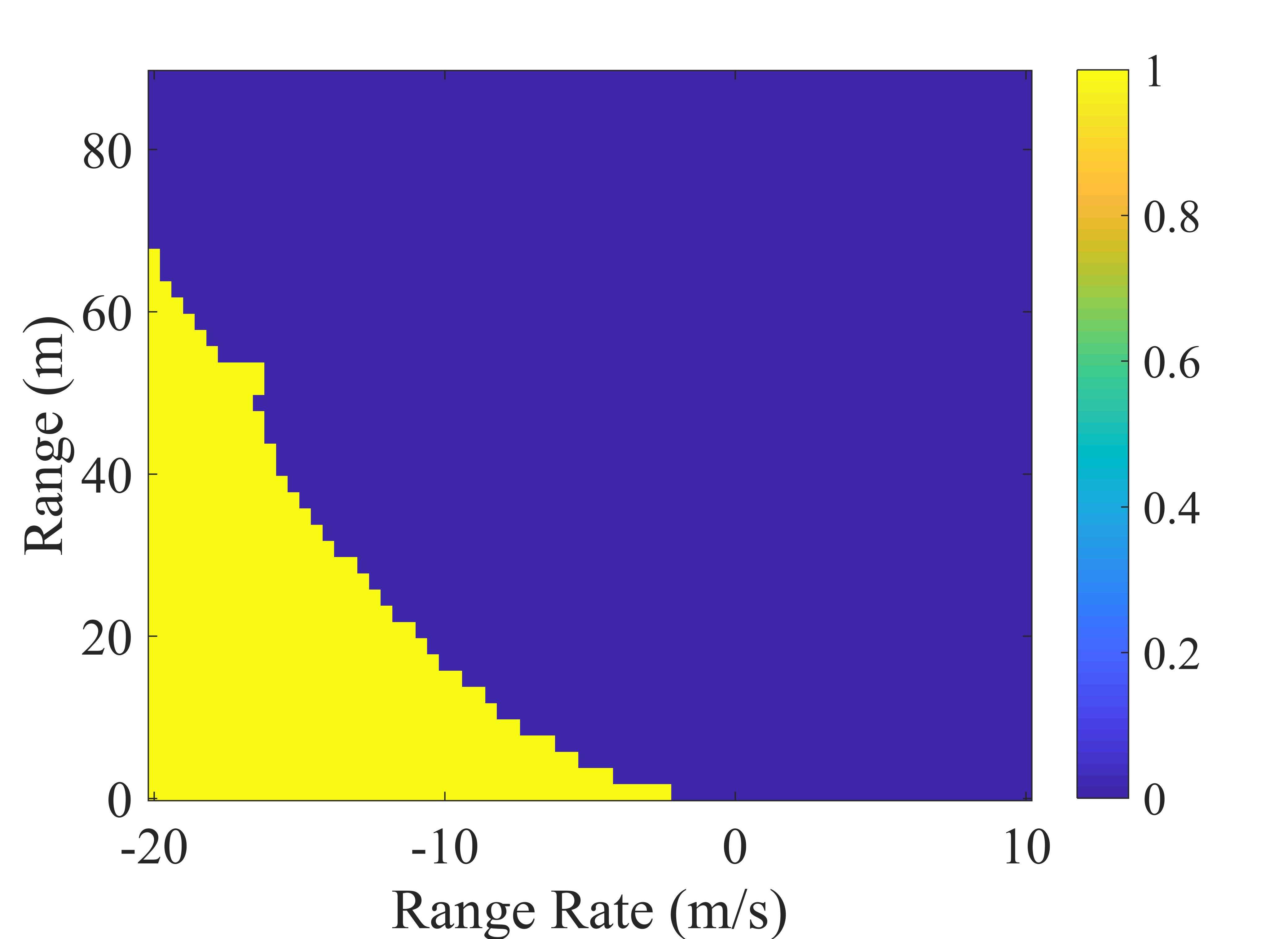}
		\centerline{\small (a) Ite-5: SM}
	\end{minipage}
	\begin{minipage}{.49\linewidth}
		\includegraphics[width=1\textwidth]{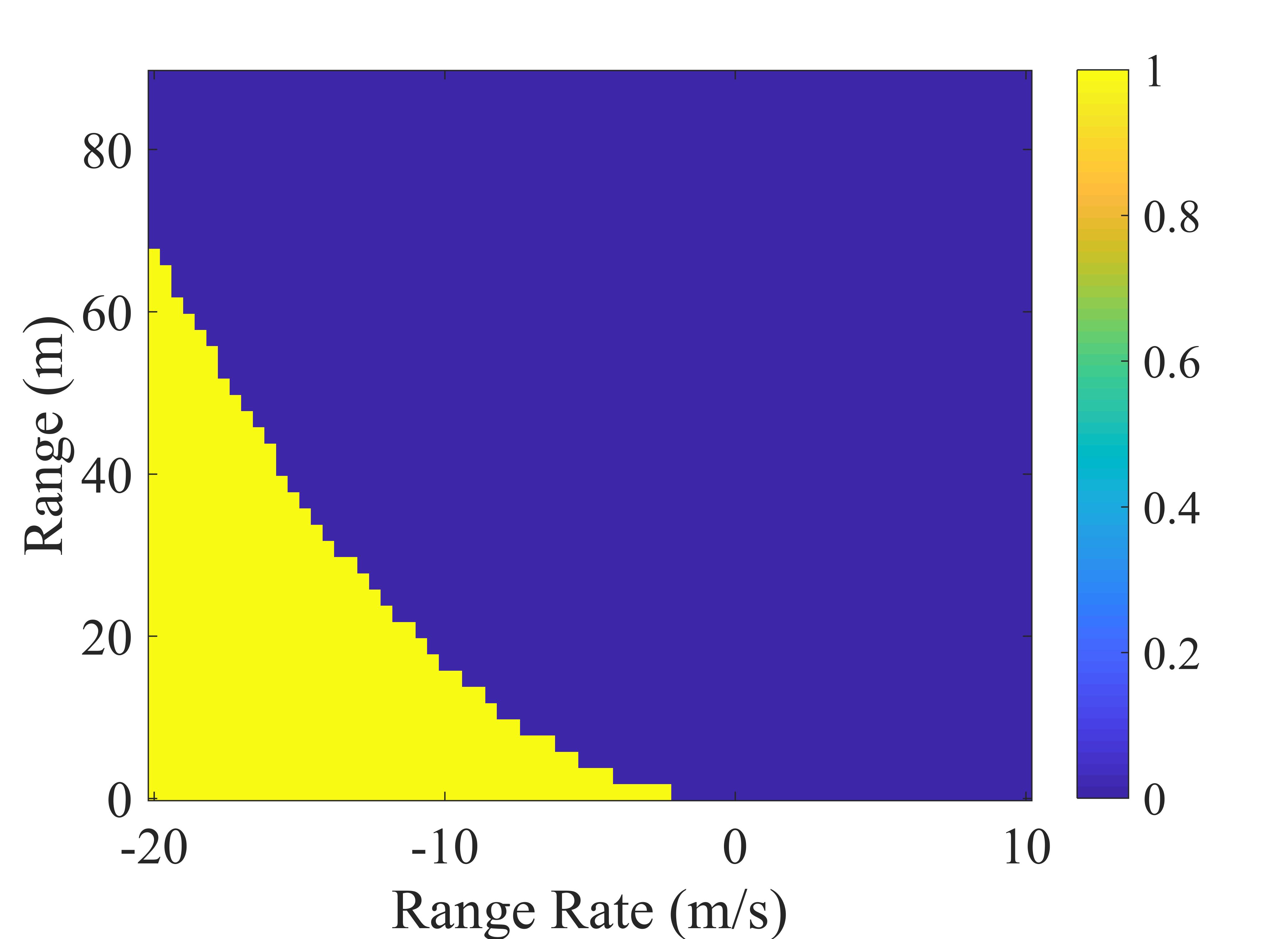}
		\centerline{\small (b) Ite-50: SM}
	\end{minipage}
	\begin{minipage}{.49\linewidth}
		\includegraphics[width=1\textwidth]{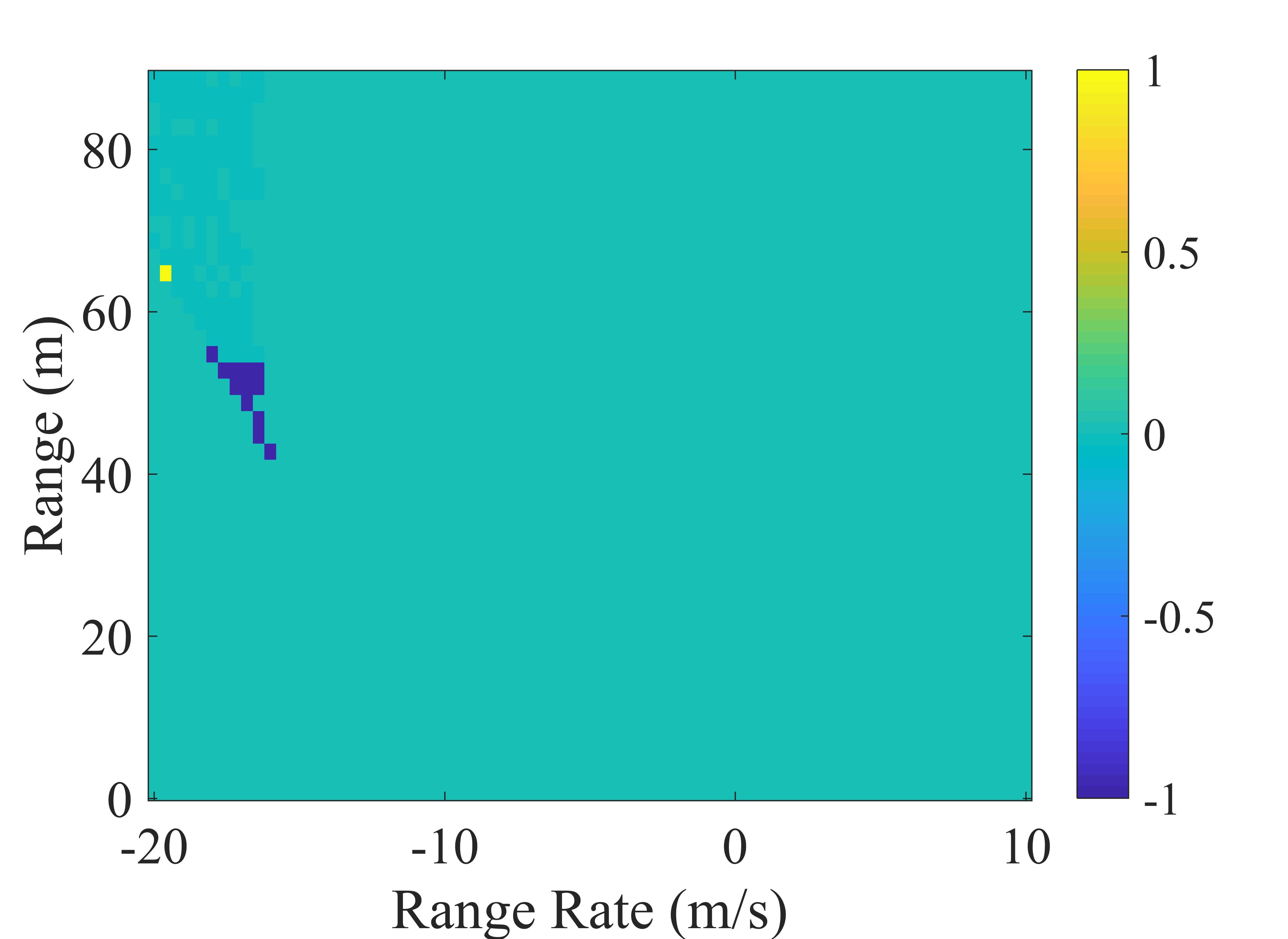}
		\centerline{\small (c) Ite-5: Dissimilarities}
	\end{minipage}
	\begin{minipage}{.49\linewidth}
		\includegraphics[width=1\textwidth]{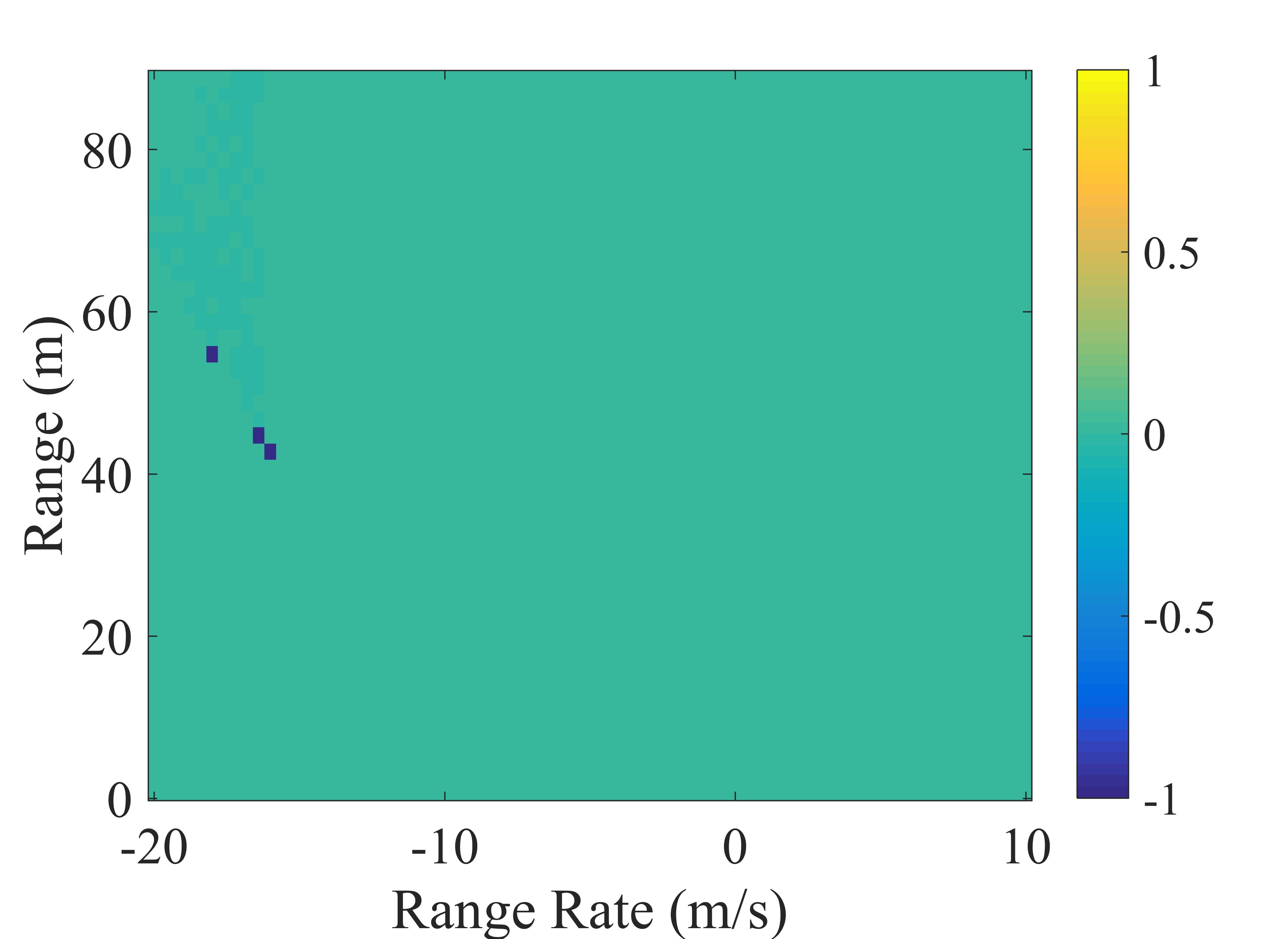}
		\centerline{\small (d) Ite-50: Dissimilarities}
	\end{minipage}
	\begin{minipage}{.49\linewidth}
		\includegraphics[width=1\textwidth]{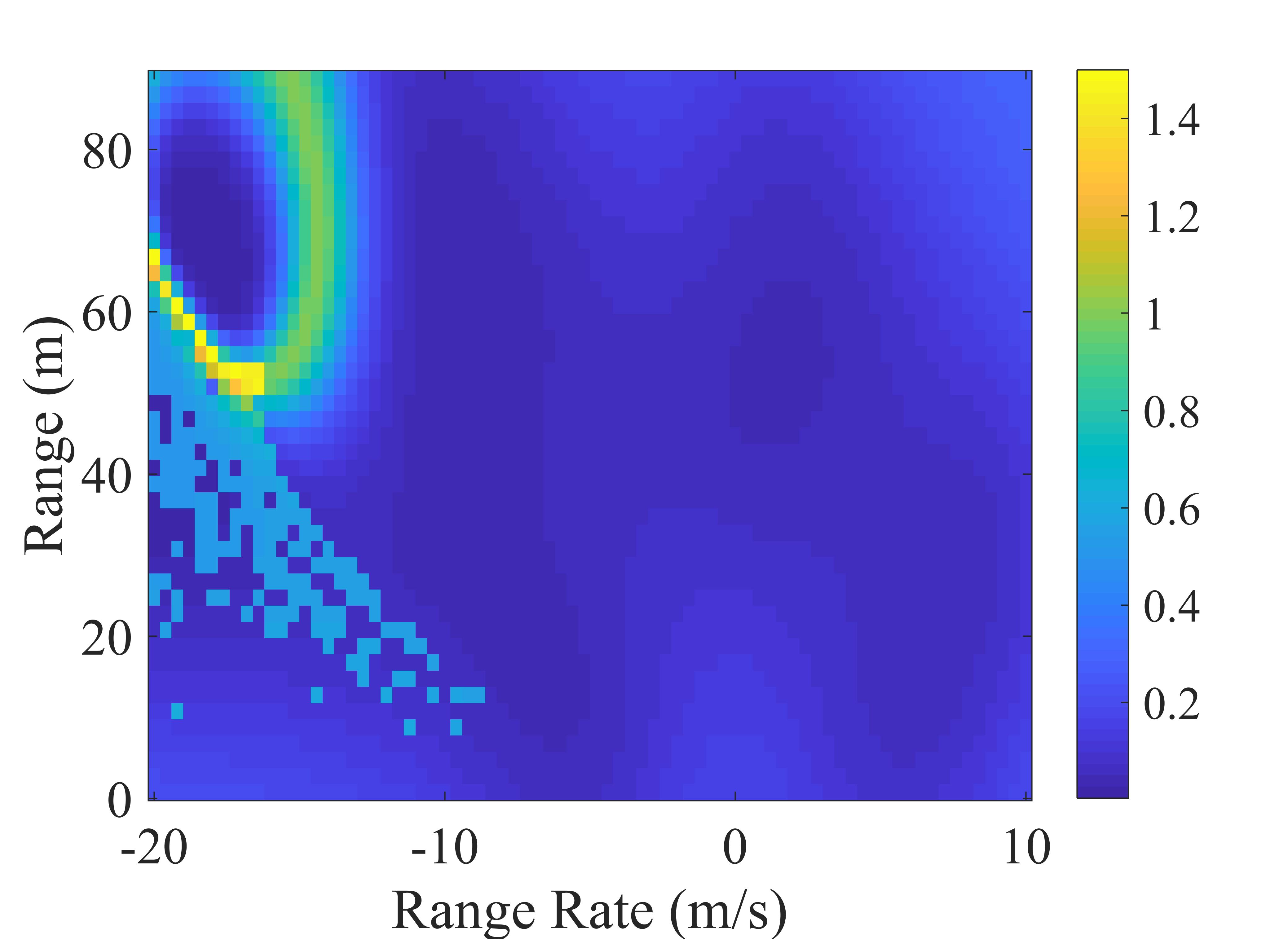}
		\centerline{\small (e) Ite-5: Acquisition Function}
	\end{minipage}
	\begin{minipage}{.49\linewidth}
		\includegraphics[width=1\textwidth]{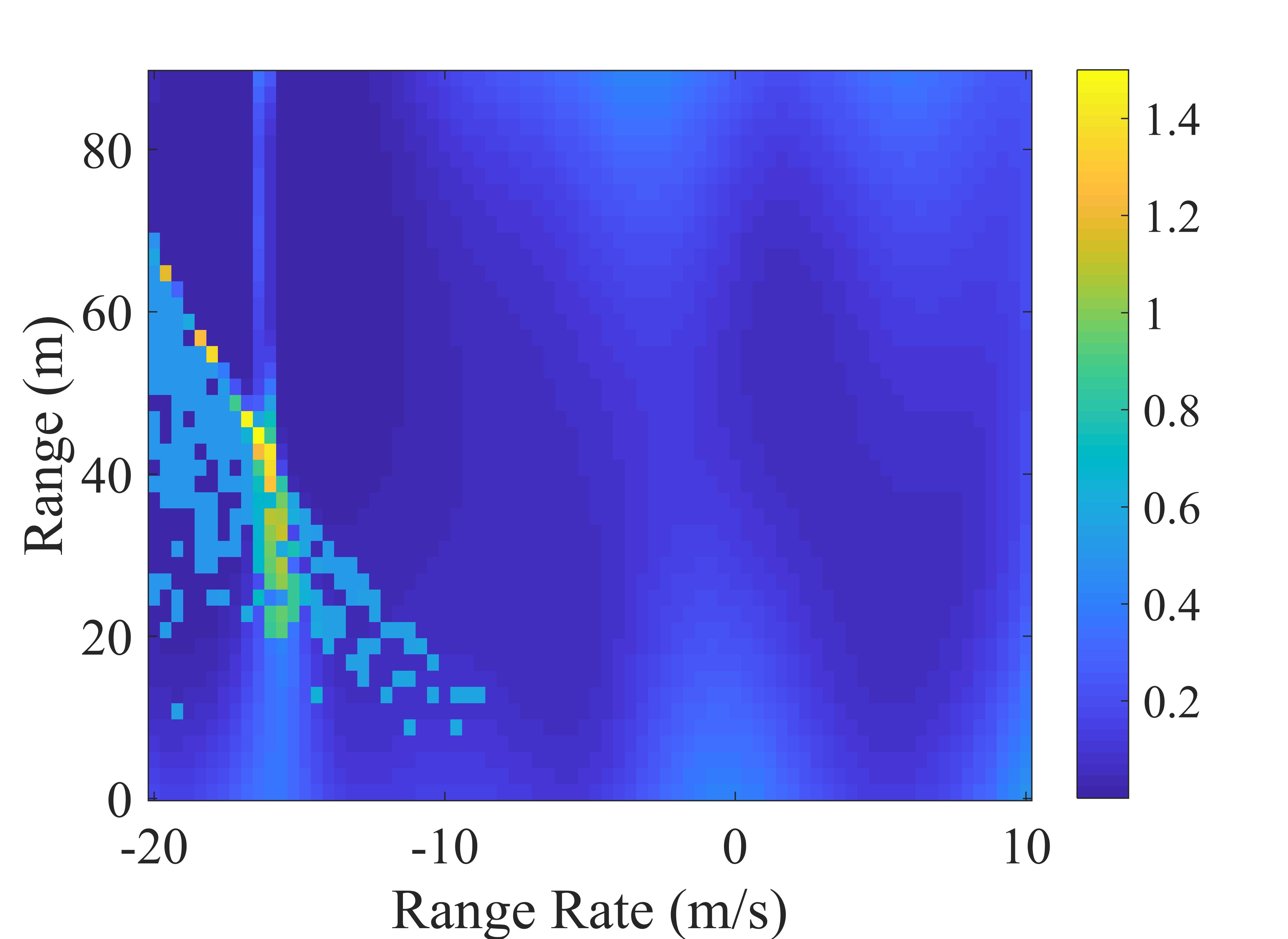}
		\centerline{\small (f) Ite-50: Acquisition Function}
	\end{minipage}
	\caption{The results of the adaptive library generation for the cut-in case. } 
	\label{fig_Cutin_Adap_Results}
\end{figure}

\begin{figure}[h!]
	\centering
	\includegraphics[width=0.35\textwidth]{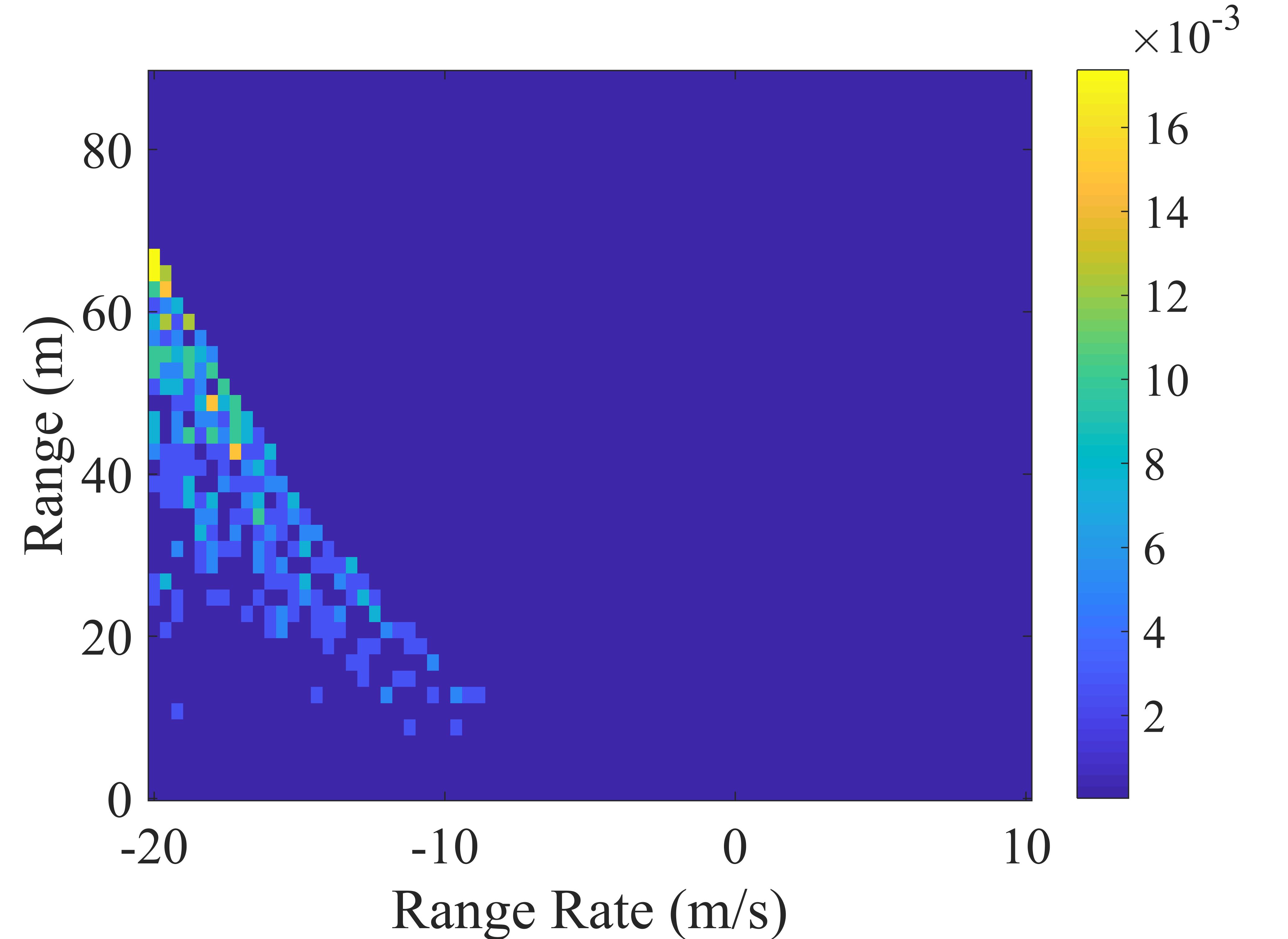}
	\caption{The customized library of the cut-in case for safety evaluation.} 
	\label{fig_ImproLib}
\end{figure}

Fig. \ref{fig_IniSam}-\ref{fig_ImproLib} show the results of the adaptive library generation process. The initial testing results are shown in Fig. \ref{fig_IniSam} (a), where the red dots denote the observed suboptimal scenarios, and the black dots denote the observed optimal scenarios. A sketch of the dissimilarity function is obtained \textcolor{black}{by the proposed method as shown in Fig. \ref{fig_IniSam} (b). To illustrate the superiority of the proposed method, the GPR method in \cite{rasmussen2003gaussian} with the same covariance function is also applied for the initial testing results as a comparison, as shown in Fig. \ref{fig_IniSam} (c). Comparing with the ground truth in Fig. \ref{fig_IniSam} (d), the GPR method is much less accurate than the proposed method. The major reason is that the dramatic change at the boundary of the dissimilarity function cannot be captured by the stationary covariance function.}

As shown in Fig. \ref{fig_Cutin_Adap_Results} (a), after 5 iterations of the adaptive testing process, performance dissimilarities between the SM and the CAV are much decreased. Fig. \ref{fig_Cutin_Adap_Results} (e) shows that the acquisition function can capture both the classification uncertainty and the regression variances. The maximization of Eq. (\ref{eq_next_sce}) is obtained by enumerating all possible scenarios $x \in \B{X}/\B{U}$. After 50 iterations, the SM has been well developed and the dissimilarities are almost eliminated, as shown in Fig. \ref{fig_Cutin_Adap_Results} (b) and (d). Comparing with the offline generated library in Fig. \ref{fig_OffLib}, the customized library has been improved significantly, as shown in Fig. \ref{fig_ImproLib}.

\subsection{CAV Evaluation}
\begin{figure}
	\centering
	\begin{minipage}{.49\linewidth}
		\includegraphics[width=1\textwidth]{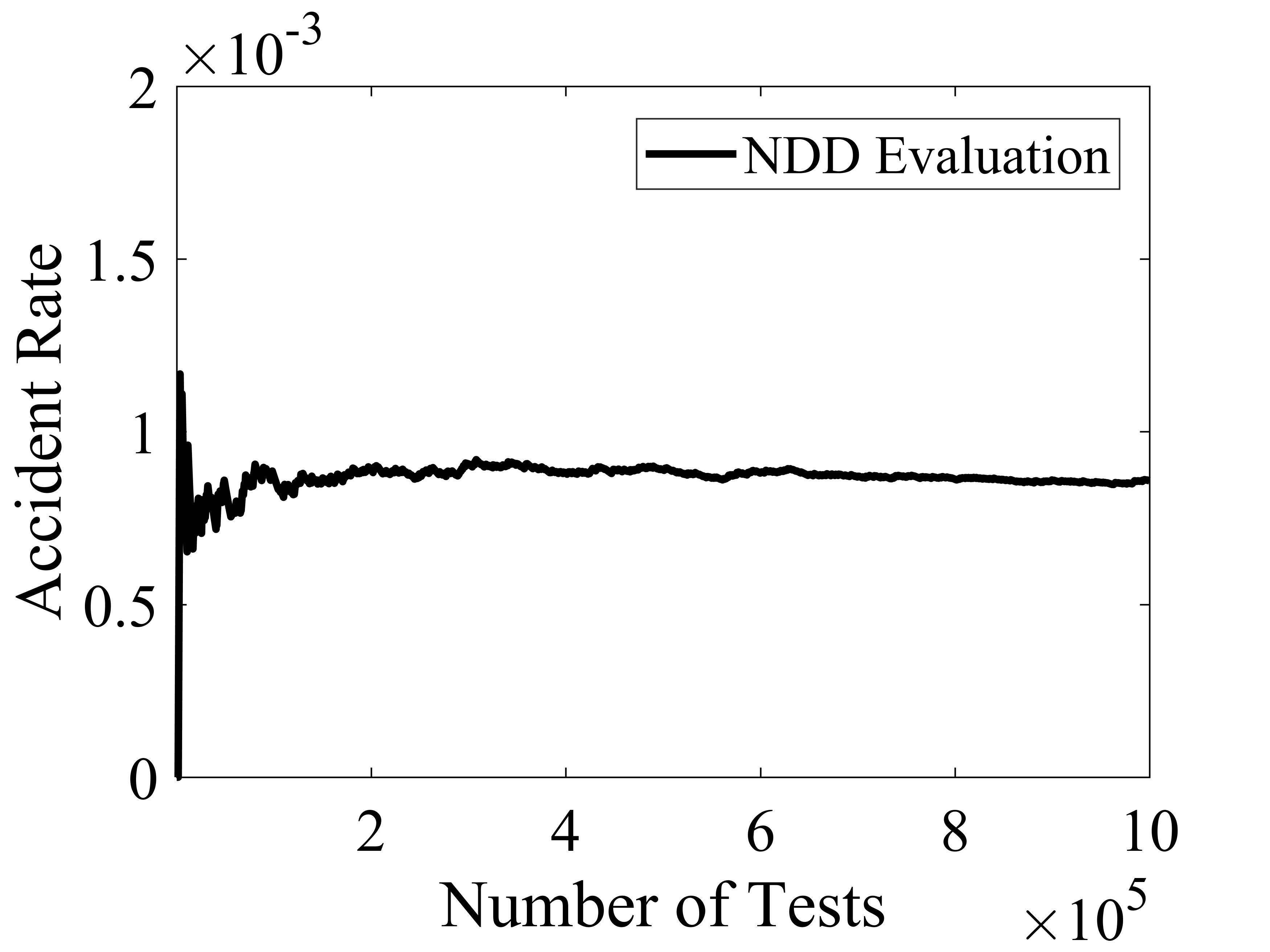}
		\centerline{(a)}
	\end{minipage}
	\begin{minipage}{.49\linewidth}
		\includegraphics[width=1\textwidth]{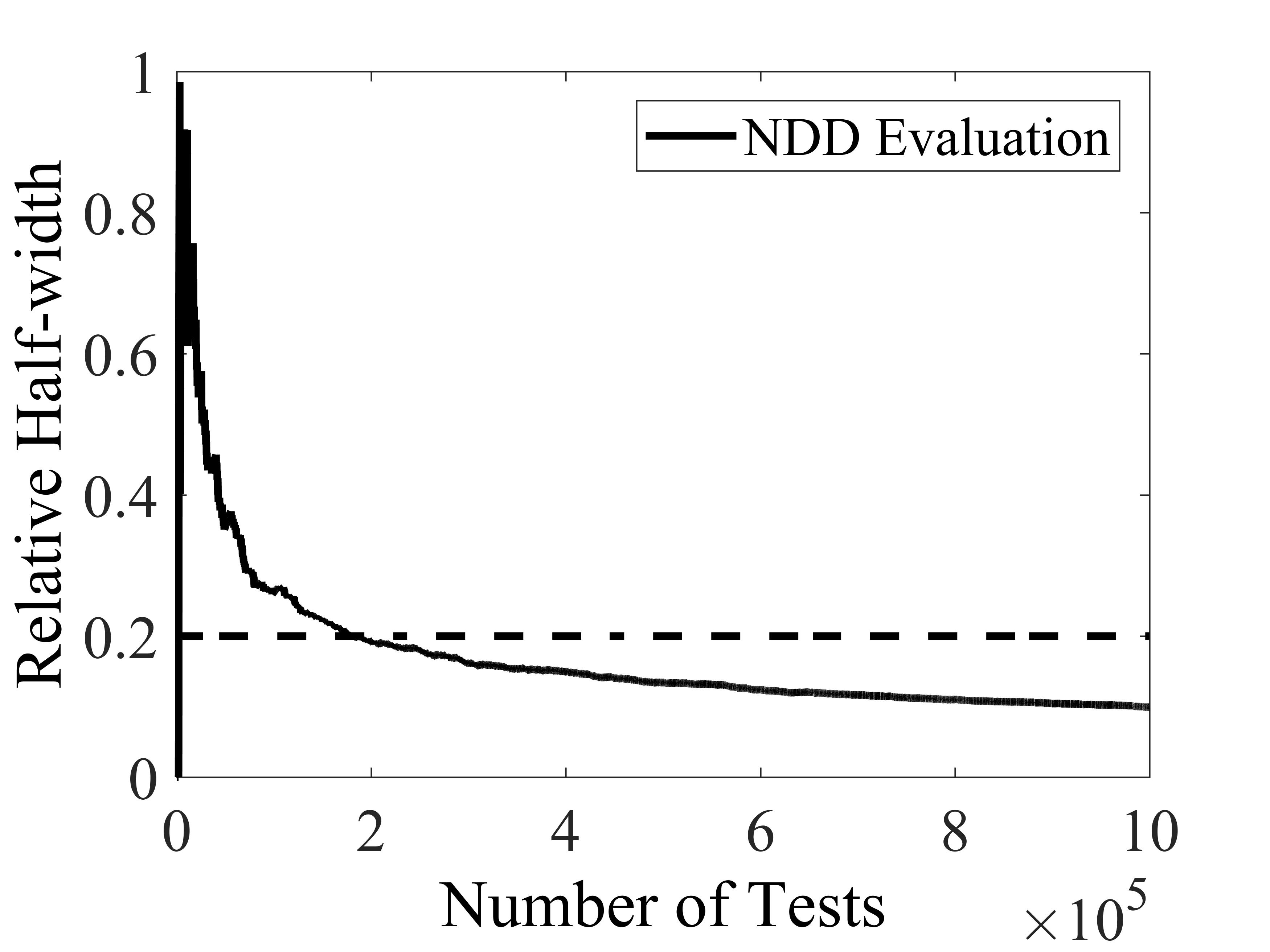}
		\centerline{(b)}
	\end{minipage}
	\begin{minipage}{.49\linewidth}
		\includegraphics[width=1\textwidth]{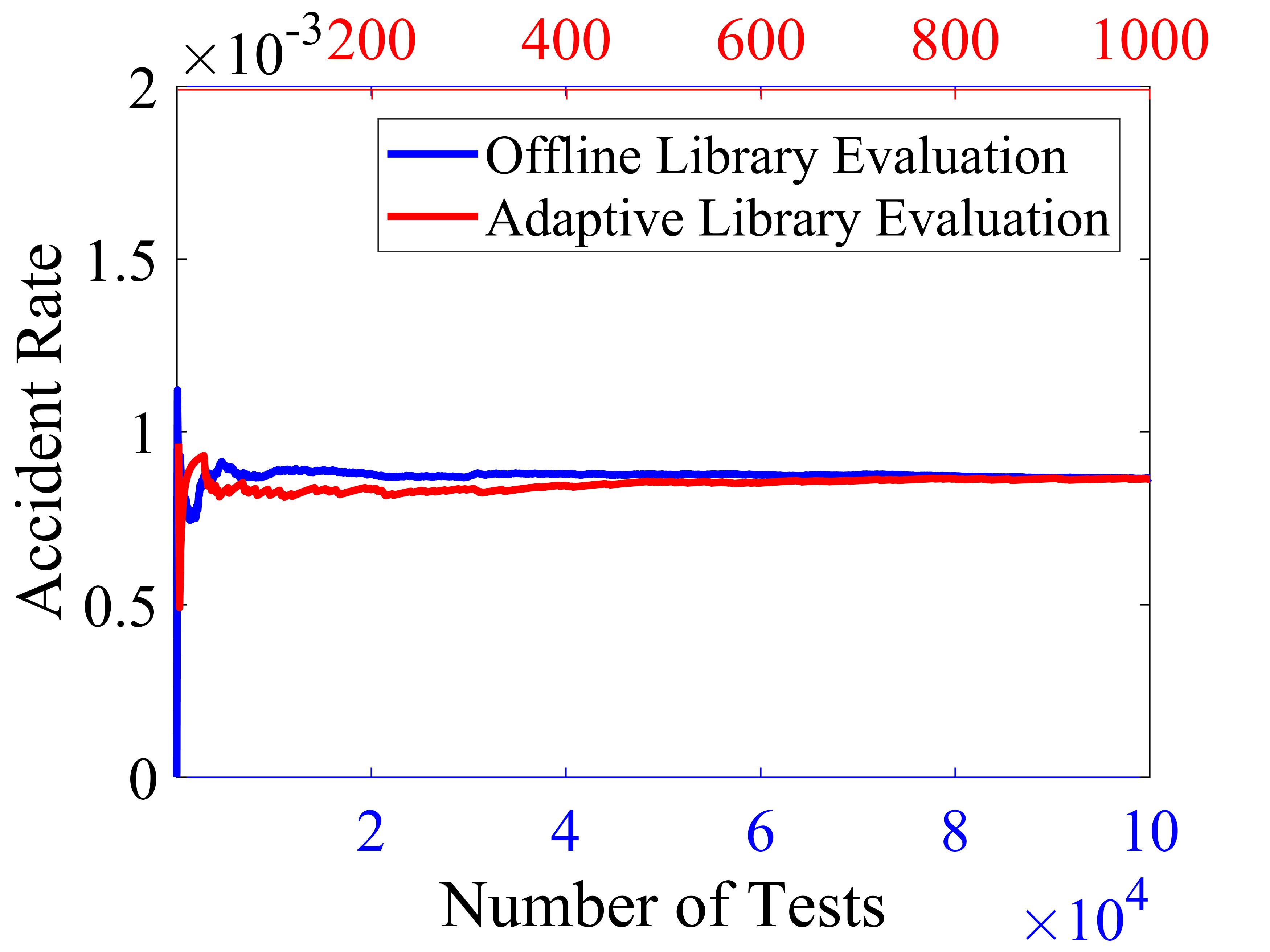}
		\centerline{(c)}
	\end{minipage}
	\begin{minipage}{.49\linewidth}
		\includegraphics[width=1\textwidth]{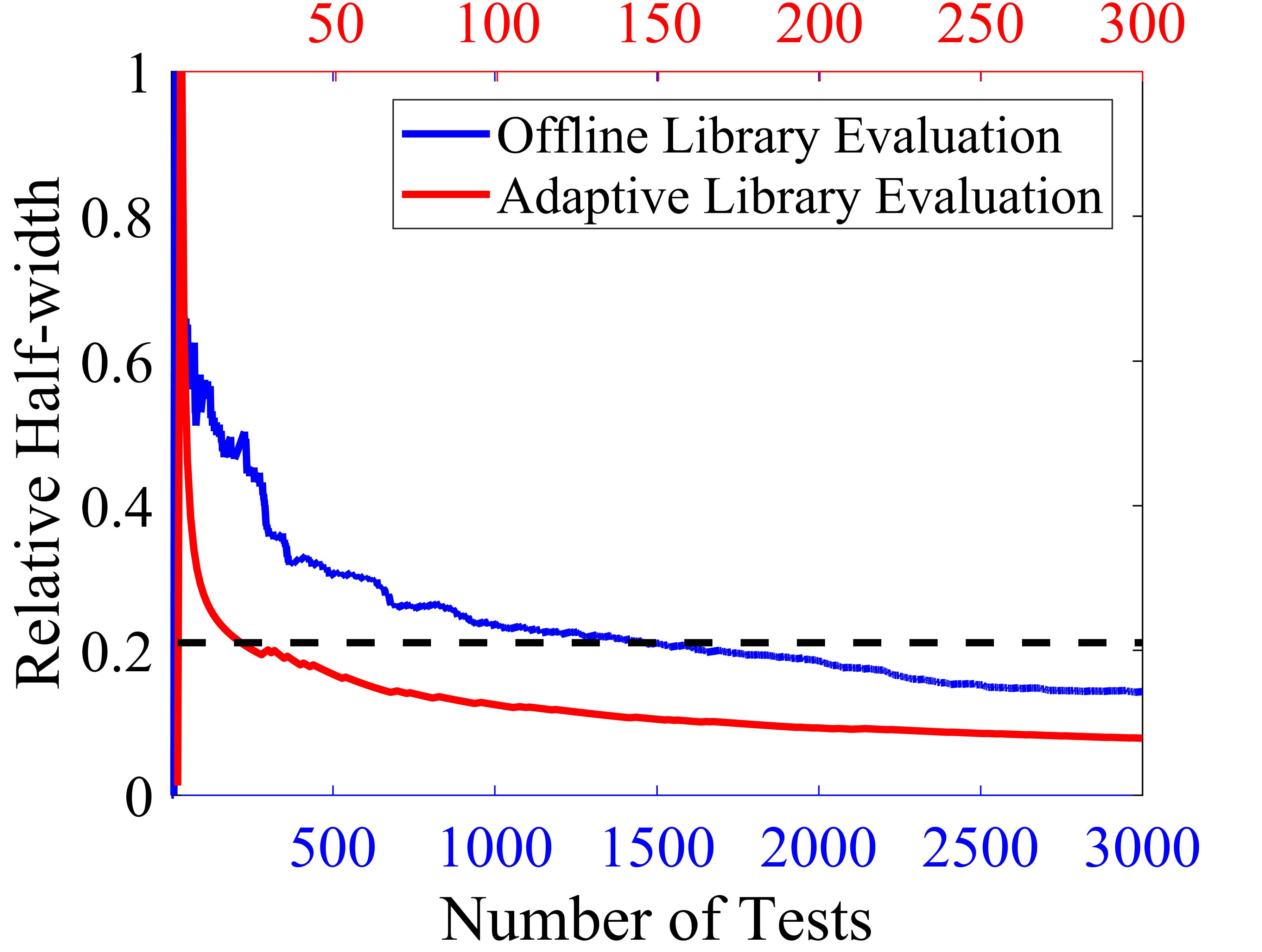}
		\centerline{(d)}
	\end{minipage}
	\caption{Results of the CAV evaluation for the cut-in case. } 
	\label{fig_Cutin_Results}
\end{figure}

\begin{figure}
	\centering
	\begin{minipage}{.49\linewidth}
		\includegraphics[width=1\textwidth]{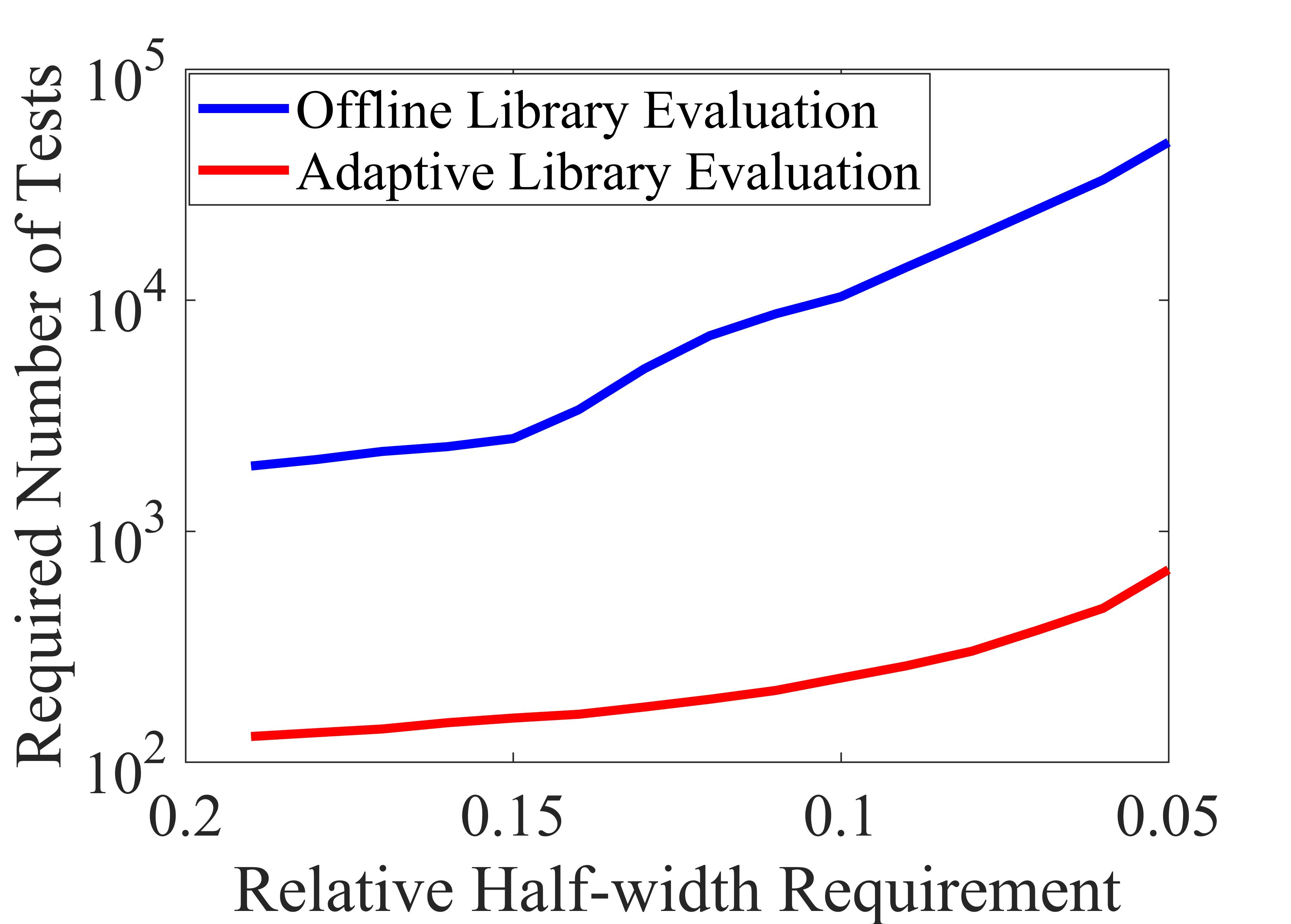}
		\centerline{(a) Required number of tests}
	\end{minipage}
	\begin{minipage}{.49\linewidth}
		\includegraphics[width=1\textwidth]{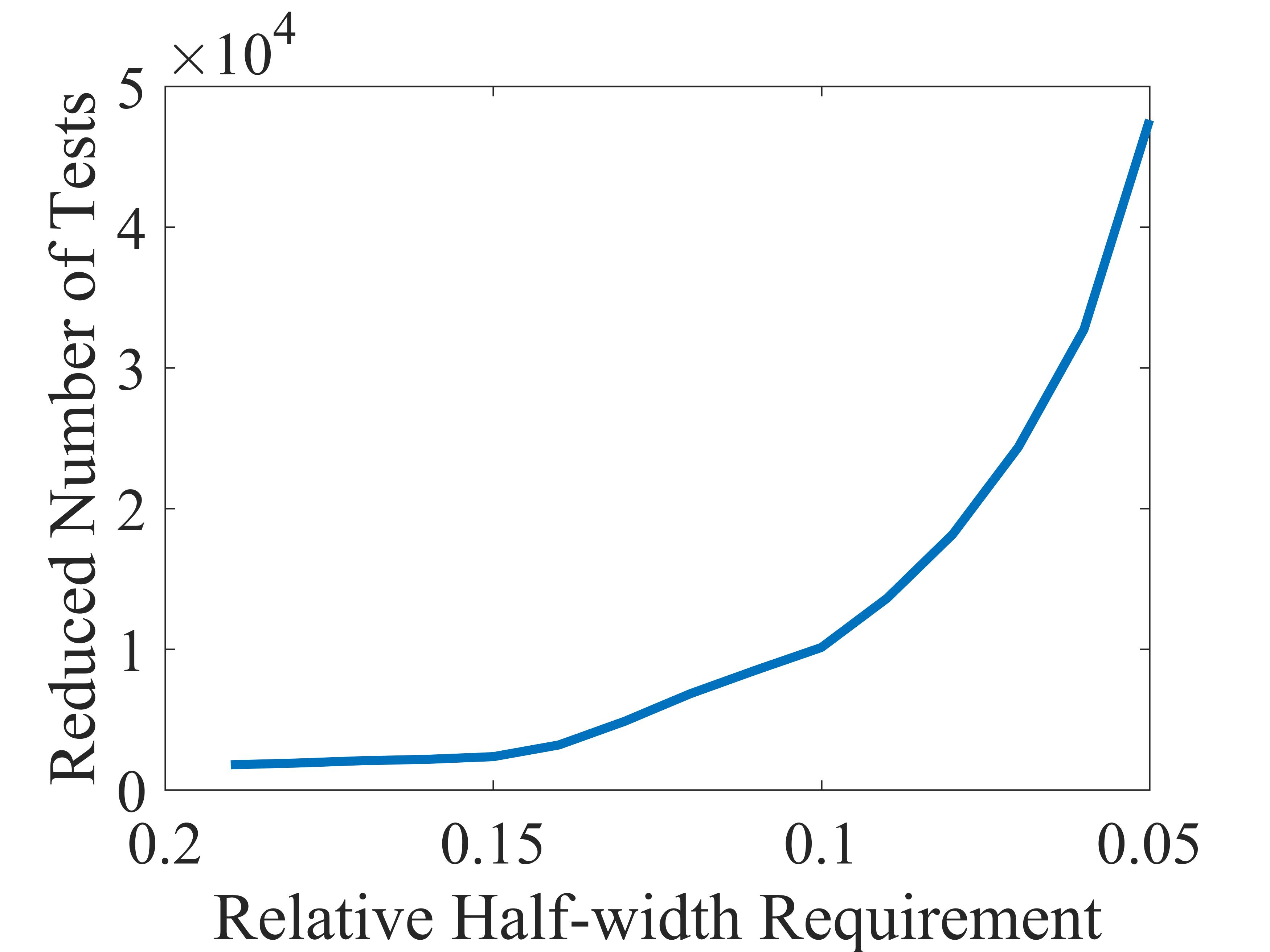}
		\centerline{(b) Reduced number of tests}
	\end{minipage}
	\caption{\color{black}The required number of tests and reduced number of tests with decreasing the required relative half-width.} 
	\label{fig_Cutin_RHW}
\end{figure}

With the customized library, the CAV is further tested and evaluated. The accident rate of the CAV is estimated by the on-road test method (i.e., NDD evaluation) and the evaluation method with the offline generated library (i.e., offline library evaluation) as two baselines. Results are shown in Fig. \ref{fig_Cutin_Results}. The blue line denotes the results of the offline library evaluation method, and the bottom $x$-axis denotes its number of tests. The red line denotes the results of the adaptive library evaluation method, and the top $x$-axis denotes its number of tests. Results show that all three methods can converge to the same accident rate after a sufficient number of tests (Fig. \ref{fig_Cutin_Results} (a) and (c)). To compare the convergence speed, the relative half-width is estimated by Eq. (\ref{eq_RHW}) with the three methods in Fig. \ref{fig_Cutin_Results} (b) and Fig. \ref{fig_Cutin_Results} (d). To reach the 0.2 relative half-width, the total required number of tests is $1.9\times10^5$, 2,090, and 121, respectively. Note that the 121 tests include 50 tests of initial scenarios, 50 tests of the adaptive testing process, and 21 tests of the CAV evaluation process. Therefore, the proposed ATSLG method accelerates the evaluation process by 1570 times and 17 times respectively, comparing with the on-road test method and the evaluation method with the offline generated library. \textcolor{black}{To further validate the reliability of the ATSLG method, we repeat the experiment 100 times and calculate the required test number for each experiment. Results show the average required test number is 121.44 and the standard deviation is 18.46. The total required numbers of tests in all 100 experiments are less than the offline library evaluation method.} Fig. \ref{fig_Cutin_RHW} shows the numbers of required tests with different required relative half-widths. By decreasing the relative half-width, the evaluation precision is increasing, and the advantage of the proposed method becomes more obvious.

\section{Conclusions}
\label{sec_conclusion}
In this paper, the adaptive testing scenario library generation (ATSLG) method is proposed to generate customized libraries for CAV testing and evaluation. Comparing with the TSLG method discussed in  \cite{feng2019testingI, feng2019testingII, feng2020safety}, the proposed method is more efficient and robust.

The major idea is to generate the customized library by compensating the dissimilarities between SM and CAV through an adaptive testing process. To leverage each test of CAV,  the Bayesian optimization scheme is applied. A classification-based Gaussian process regression is adopted to estimate the non-stationary dissimilarity function, and a new acquisition function is designed to determine new testing scenarios in each iteration. A cut-in case is investigated for safety evaluation. Comparing with the TSLG method, the total number of required tests is further decreased by a few orders of magnitude (e.g., 10-100 times). More importantly, the acceleration of the evaluation process is more prominent if higher precision is required.

There are still many interesting topics that can be further investigated. For example, when the ATSLG problem in high-dimensional scenarios becomes more complex, how to address the high-dimensional issue in the adaptive process remains as a problem. Moreover, it is interesting to apply the proposed method in more realistic CAV testing platforms with pre-established scenario libraries.

\appendices
\appendices

\bibliographystyle{IEEEtran}
\bibliography{TSLG-III-07162020}

\begin{IEEEbiography}[{\includegraphics[width=1in,height=1.25in,clip,keepaspectratio]{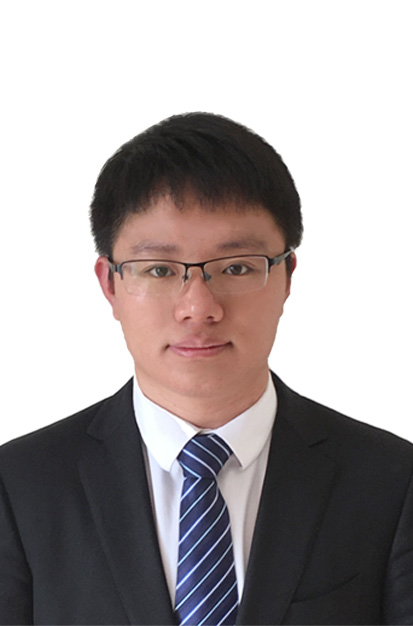}}]{Shuo Feng}
	is currently a postdoctoral researcher at Department of Civil and Environmental Engineering at University of Michigan, Ann Arbor. He received the B.S. and Ph.D. degrees in Department of Automation from Tsinghua University, China, in 2014 and 2019, respectively.  He was also a joint Ph.D. student at Department of Civil and Environmental Engineering at University of Michigan, Ann Arbor, in 2017-2019. His current research interests include testing, evaluation, and optimization of connected and automated vehicles.
\end{IEEEbiography}
\vspace{-10 mm}

\begin{IEEEbiography}[{\includegraphics[width=1in,height=1.25in,clip,keepaspectratio]{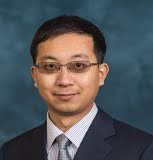}}]{Yiheng Feng}
	is currently an Assistant Research Scientist at University of Michigan Transportation Research Institute. He graduated from the University of Arizona with a Ph.D degree in Systems and Industrial Engineering in 2015. He has a Master degree from the Civil Engineering Department, University of Minnesota, Twin Cities in 2011. He also earned the B.S. and M.E. degree from the Department of Control Science and Engineering, Zhejiang University, Hangzhou, China in 2005 and 2007 respectively. His research interests include traffic signal systems control and security, and connected and automated vehicles testing and evaluation.
\end{IEEEbiography}
\vspace{-10 mm}
\begin{IEEEbiography}[{\includegraphics[width=1in,height=1.25in,clip,keepaspectratio]{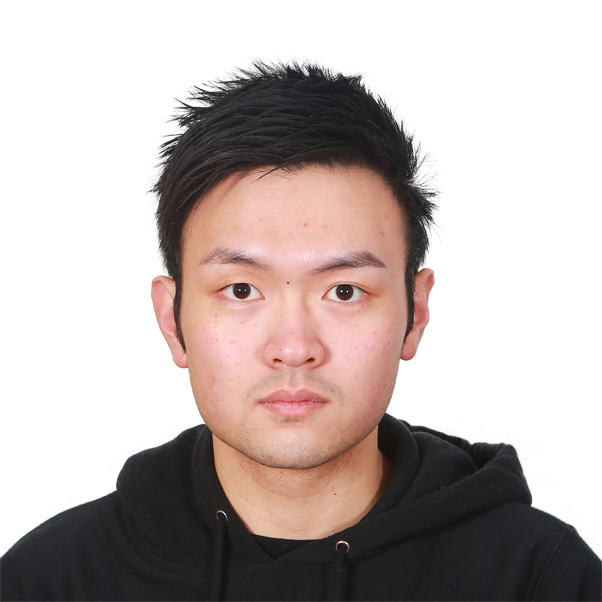}}]{Haowei Sun}
	is currently a graduate at Department of Civil and Environmental Engineering at University of Michigan. He received the bachelor’s degree in Department of Automation  from Tsinghua University, China, in 2019, and he visited University of Michigan for a summer research internship in 2018. His research interests include intelligent transportation, optimization method and deep reinforcement learning.
\end{IEEEbiography}
\vspace{-10 mm}
\begin{IEEEbiography}[{\includegraphics[width=1in,height=1.25in,clip,keepaspectratio]{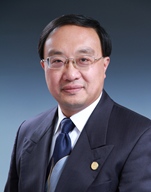}}]{Yi  Zhang}
	received   the   BS   degree   in 1986   and   MS   degree   in   1988  from Tsinghua University in China, and earned  the  Ph.D.  degree  in  1995  from  the University  of  Strathclyde  in  UK.  He  is a   professor   in   the   control   science   and engineering  at  Tsinghua  University  with his  current  research  interests  focusing  on intelligent  transportation  systems. His  active  research  areas include  intelligent  vehicle-infrastructure  cooperative  systems, analysis  of  urban  transportation  systems,  urban  road  network management,  traffic  data  fusion  and  dissemination,  and  urban traffic control and management.   His research fields also cover the advanced control theory and applications, advanced detection and measurement, systems engineering, etc.
\end{IEEEbiography}
\vspace{-10 mm}
\begin{IEEEbiography}[{\includegraphics[width=1in,height=1.25in,clip,keepaspectratio]{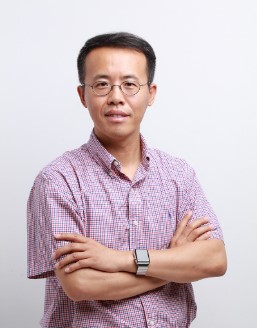}}]{Henry X. Liu}
	is a Professor of Civil and Environmental Engineering at the University of Michigan, Ann Arbor and a Research Professor of the University of Michigan Transportation Research Institute. He also directs the USDOT Region 5 Center for Connected and Automated Transportation. Dr. Liu received his Ph.D. degree in Civil and Environmental Engineering from the University of Wisconsin at Madison in 2000 and his Bachelor degree in Automotive Engineering from Tsinghua University in 1993. Dr. Liu's research interests focus on transportation network monitoring, modeling, and control, as well as mobility and safety applications with connected and automated vehicles. On these topics, he has published more than 100 refereed journal articles. Dr. Liu is the managing editor of Journal of Intelligent Transportation Systems and an associate editor of Transportation Research Part C. 
\end{IEEEbiography}

\end{document}